\documentclass[pdflatex,sn-mathphys]{sn-jnl}

\usepackage{physics} 
\usepackage[normalem]{ulem} 
\usepackage{bbm}
\usepackage{graphicx}
\usepackage{bm}
\usepackage[T1]{fontenc} 
\usepackage{amsmath}
\usepackage{amssymb}
\usepackage{amsthm}

\theoremstyle{thmstyleone}

\theoremstyle{thmstyletwo}

\theoremstyle{thmstylethree}

\renewcommand{\rm}{\mathrm}
\renewcommand{\d}{\,\text{d}}

\begin{document}

\title{Can the dipolar interaction suppress dipolar relaxation?}

\author*[1]{\fnm{Pierre} \sur{Barral}}\email{pbarral@mit.edu}
\equalcont{These authors contributed equally to this work.}

\author[1]{\fnm{Michael} \sur{Cantara}}
\equalcont{These authors contributed equally to this work.}

\author[1]{\fnm{Li} \sur{Du}}
\equalcont{These authors contributed equally to this work.}

\author[1]{\fnm{William} \sur{Lunden}}

\author[1]{\fnm{Julius} \sur{de Hond}}

\author[1]{\fnm{Alan O.} \sur{Jamison}}

\author[1]{\fnm{Wolfgang} \sur{Ketterle}}

\affil[1]{Research Laboratory of Electronics, MIT-Harvard Center for Ultracold Atoms, and Department of Physics, Massachusetts Institute of Technology, Cambridge, Massachusetts 02139, USA}

\maketitle


\textbf{Magnetic atoms in a thin layer have repulsive interactions when their magnetic moments are aligned perpendicular to the layer. We show experimentally and theoretically how this can suppress dipolar relaxation, the dominant loss process in spin mixtures of highly magnetic atoms. Using dysprosium, we observe an order of magnitude extension of the lifetime, and another factor of ten is within reach based on the models which we have validated with our experimental study. The loss suppression opens up many new possibilities for quantum simulations with spin mixtures of highly magnetic atoms.}

\section*{Introduction}
Experiments with ultracold atoms or molecules are often limited by unfavorable inelastic collision rates. Several methods have been developed to control collisions such as isolating atoms in deep lattices \cite{wilpers2002optical}, reducing collisional channels via confinement \cite{Pasquiou10}, or by mitigating their effects through the enhancement of elastic collisions via Feshbach resonances \cite{Cornish2000}. Polar molecules, in particular, have been shielded from chemical reactions at short range by using repulsive interactions between electric dipoles, either in two dimensions or via microwave dressing \cite{valtolina2020dipolar, Anderegg2021}.

Here we explore how dipolar shielding can be realized in dysprosium, a highly magnetic atom for which the dipolar interaction is two orders of magnitude smaller than for polar molecules. Magnetic atoms have a simpler structure than molecules, allowing them to achieve lower temperatures while providing a controlled, tunable, and relatively simple platform for exploring novel forms of matter with long-range forces \cite{chomaz2022dipolar, lahaye2009physics, Lian2012, Deng2012, Lian2014, Cui2013, Babik20}. Dysprosium, with a magnetic moment of $10~\mu_B$, has a magnetic dipole-dipole interaction that is 100 times larger than that of alkali atoms. However, dipolar relaxation -- an inelastic spin-flip process that converts Zeeman energy into kinetic energy -- occurs at a rate that also scales as the square of the dipolar interaction, severely limiting the lifetime of any cloud with population in an excited Zeeman level. The dominance of dipolar relaxation has, thus far, precluded the experimental realization of many proposed new phenomena in spin mixtures of highly magnetic atoms \cite{Gopalakrishnan2013, Yi2006, Lev15}.

Using dipolar shielding to prevent the atoms from undergoing dipolar relaxation requires a deep understanding of the dipolar interaction as it drives both the elastic and inelastic processes. Nevertheless, as we show here, suppression of dipolar relaxation is possible since it occurs mainly at specific interatomic separations, where the dipolar potential possibly reduces the wave function amplitude. We have observed an order of magnitude suppression of the dipolar relaxation rate, and, supported by comprehensive simulations of the decay rate, we show that another order of magnitude is within reach given reasonable parameters. In the limit of high magnetic fields, or for very low temperatures, the amount of suppression can be made arbitrary large. We first describe qualitatively the interplay of magnetic field, temperature, and shielding, then present our experimental results, followed by theoretical simulations.

\section*{Basic principles}
The dipole-dipole interaction is attractive in the case of a tip-to-tail orientation and repulsive for the side-by-side one. Constraining atoms to an $xy$ plane, with a magnetic moment aligned perpendicularly along $z$, leads to a largely side-by-side repulsion and generates a dipolar barrier. The dipolar length\footnote{Here we use the definition used broadly for 2-body collisions. In a many-body physics context, the alternative definition $a_\rm{dd} = \frac{\mu_0}{4\pi}\frac{2\mu(10\mu_B)^2}{3\hbar^2}$ is more common.} $a_\rm{dd} = \frac{\mu_0}{4\pi}\frac{\mu(10\mu_B)^2}{\hbar^2}$ represents the strength of the interaction and the two-particle oscillator length $a_z = \sqrt{\hbar/\mu\omega_z}$ the extension of the cloud in the $z$ direction. We denoted $\mu$ and $\omega_z$ the reduced mass and the trap frequency respectively. A dipolar barrier appears when the dipolar length $a_\rm{dd} > 0.34~a_z$ \cite{Ticknor10}, which we refer to as the quasi-2D regime. Thus, experiments with dysprosium require 10,000 times higher axial frequencies than polar molecules to compensate for the 100 times smaller dipolar length. Our experiments have reached this regime with $a_z = 20$~nm and $a_\rm{dd} = 10$~nm.

\begin{figure}[h]
    \centering
    \includegraphics[width=0.6\columnwidth]{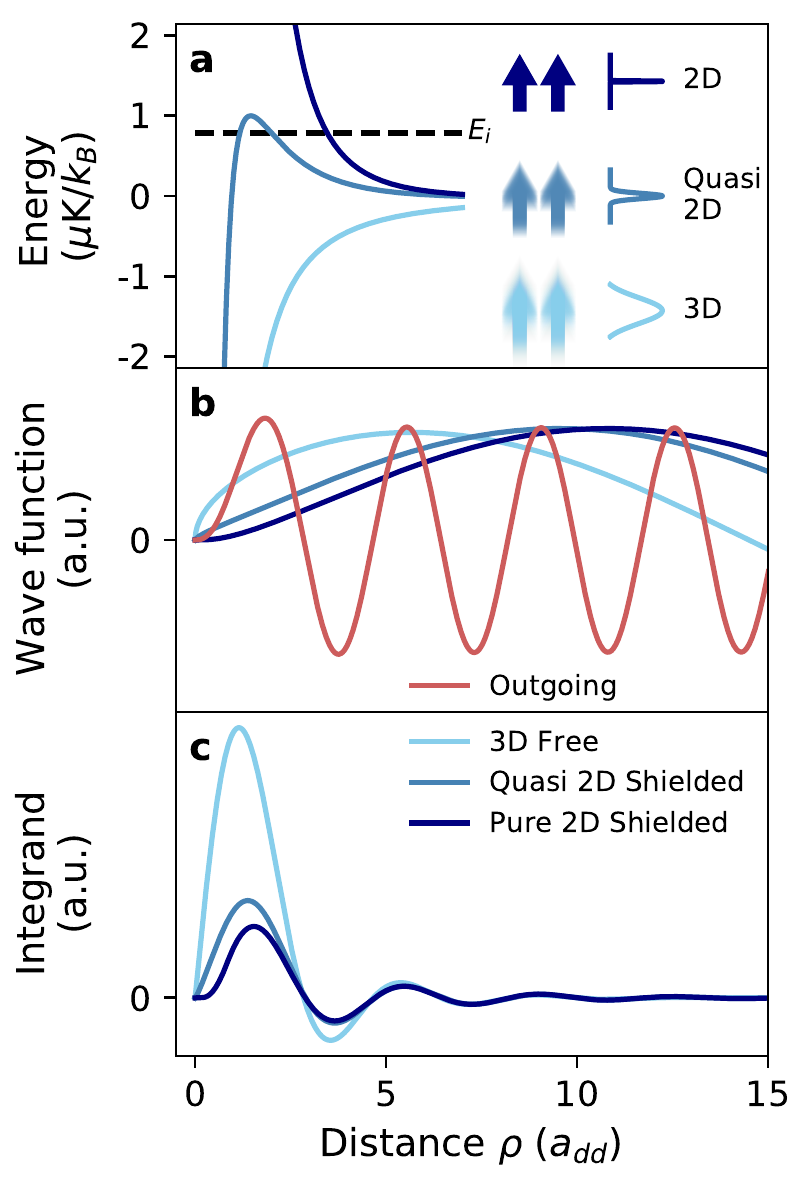}
    \caption{\textbf{Principle of dipolar shielding}  (a.) Effective radial potential between two atoms from equation~(\ref{eq:differentialEquation}) for: no confinement (light blue, 3D), quasi-2D with $\omega_z/2\pi = 300~\mathrm{kHz}$ (steel blue) and pure-2D (dark blue). The incoming energy is given by the temperature $T =~1~\mu$K. (b.) Wave function solutions of equation~(\ref{eq:differentialEquation}) with $n=0$ and initial orbital momentum $m_i=0$ for the three confinement strengths described above, and in red the spin-flipped outgoing wave function to $m_f=2$ and $B=500$~mG. The effect of shielding of the outgoing wave function is negligible for these parameters. (c.) Integrand of Fermi's golden rule (equation~(\ref{eq:FermiGoldenRule}) and see equation~(\ref{eq:3DdecayRate}) in Supplementary Information). Each curve is the product of the respective wave function in (b.), the outgoing wave function, and the double spin-flip operator from equation~(\ref{eq:dipoleOperatorOnState}) integrated with the harmonic oscillator wave functions in the $z$-direction. The shielding we implement here corresponds to the difference between the light blue and steel blue curves. The minimum attainable decay rate for this incoming energy corresponds to the dark blue curve. See Supplementary Information for insights on the behavior of the integrand.}  
    \label{fig:plotWavefunctions_singleChannel}
  \end{figure}

Three parameters determine the loss rate in quasi-2D: the ratio $a_\rm{dd}/a_z$ set by the confinement, the temperature $T$, and the magnetic field $B$. The potential barrier increases with confinement, ultimately reaching the pure-2D limit as $a_z \rightarrow 0$ as shown Fig.~\ref{fig:plotWavefunctions_singleChannel}a. As the temperature decreases, the wave function of an incoming pair is suppressed by the barrier over a longer range, thereby decreasing the chance of two atoms reaching close range. This shielding effect on the wave function is illustrated in Fig.~\ref{fig:plotWavefunctions_singleChannel}b. As the magnetic field increases, the range where dipolar relaxation occurs is shortened and the shielding increases. Indeed, a higher magnetic field leads to a higher released energy, and correspondingly a more rapidly oscillating outgoing wave function (see red curve Fig.~\ref{fig:plotWavefunctions_singleChannel}b). Since the dipolar potential falls off as $1/r^3$, the majority of the decay will come from the first oscillating lobe of the outgoing wave function, as seen in Fig.~\ref{fig:plotWavefunctions_singleChannel}c. The range of dipolar relaxation, therefore, decreases as the magnetic field increases. This can also be explained in a semi-classical picture: the Franck-Condon principle predicts spin flips to occur at the classical turning point of the outgoing wave function \cite{Condon47, Pasquiou10}, i.e.\ when the released Zeeman energy equals the energy of the centrifugal barrier. Correspondingly, higher magnetic fields cause spin-flips to occur at a shorter range, ultimately behind the barrier felt by the incoming atoms, where they are strongly suppressed. Therefore, shielding qualitatively changes the magnetic field dependence of the dipolar relaxation rate. For bosons, in both 3D \cite{Hensler03, Pasquiou10, Lev15} and unshielded 2D geometries, the relaxation rate increases with magnetic field. When accounting for the barrier in 2D, however, the signature of dipolar shielding appears: the relaxation rate decreases with magnetic field (see Supplementary Information).

\section*{Experiment}
\begin{figure}[h]
    \centering
    \includegraphics[width=1\columnwidth]{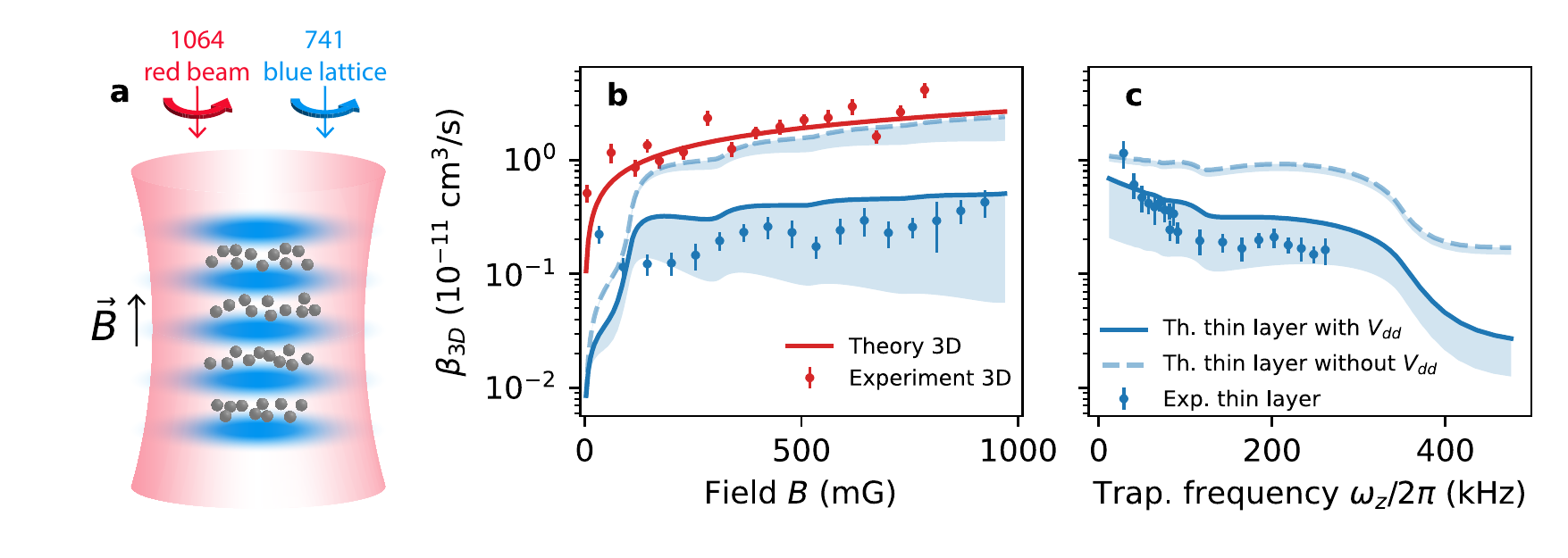}
    \caption{
    \textbf{Experiment scheme and results.} (a.) Trap geometry. A blue-detuned $741~\mathrm{nm}$ retroreflected beam repels the atoms to create a 1D lattice. The finite contrast of the lattice and the zero-point motion of the atoms in the ground state create a repulsive transverse potential, which is compensated by a $1064~\mathrm{nm}$ red-detuned beam to create an adjustable transverse harmonic confinement.
    (b. and c.) Experimentally measured $\beta_\mathrm{3D}$ in a large volume trap (red) and in a thin layer (blue). The lines are theory curves obtained by using Fermi's golden rule (see Supplementary Information for derivations).  The red curve shows the decay rate in 3D \cite{Hensler03, Lev15}, the dashed blue curve is  for non-shielded atoms in a lattice \cite{Pasquiou10}.  The solid blue line takes into account the shielding induced by the elastic dipole-dipole interaction. All theoretical curves are thermally averaged over the incoming momenta. The shaded blue region corresponds to the inclusion of van der Waals contact interactions (see Supplementary Information). (b.) Measurement of $\beta_\mathrm{3D}$ as a function of magnetic field. The axial trap frequency is $\omega_z/2\pi = 185~\mathrm{kHz}$ which corresponds to $a_z/2 = 13~\mathrm{nm}$. (c.) Measurement of $\beta_\rm{3D}$ in a constant magnetic field of $200~\mathrm{mG}$ while varying the trap frequency. The uncertainties are set by the atom number stability, cloud temperature measurement and trap frequency measurements (see Methods section).}
    \label{fig:results}
\end{figure}

Here we study these principles experimentally. We load $\sim 8\times 10^4$ spin-polarized $^{162}\rm{Dy}$ atoms in the excited $\ket{J = 8, m_J = 8}$ Zeeman level (see Methods for details) in an optical lattice and get a stack of about 45 thin pancakes (`crêpes'). The crêpes reach an $a_z/2 = 10$~nm root-mean-square (RMS) width and a 5.7~$\mu$m radius. The peak density is $2.9\times 10^9\,\rm{cm}^{-2}$. The experiment is performed at $T\approx 1.6~\mu$K, above the BEC transition temperature (300~nK), to prevent convolving our results with changes in the two-particle correlation function \cite{Kagan85, Burt97}. The quantization axis is set by an external magnetic field along the $z$ direction. The lattice beam is blue detuned, with its radial repulsion compensated by a coaxial red-detuned optical dipole trap, as shown in Fig.~\ref{fig:results}a. Axial trap frequencies are limited to $\omega_z/2\pi = $~260~kHz by the maximum laser power of the compensation beam.

By measuring the atom losses we determine the inelastic decay coefficient, $\beta_\mathrm{3D}$, as defined by the differential equation for the 3D density $n$:
\begin{equation}
\label{eq:two-body-decay-diff_eq}
    \frac{dn}{dt} = -\beta_\mathrm{3D}n^2.
\end{equation}
We obtain densities from the measured atom number, temperature and trap frequencies, and average over the stack of crêpes (also see the Methods section).
We sometimes refer to the 2D loss rate $\beta_\mathrm{2D}$ in $\mathrm{cm^2/s}$, which uses the 2D density instead. It is related to $\beta_\mathrm{3D}$ through the axial harmonic confinement via $\beta_\mathrm{2D} = \beta_\mathrm{3D}/(a_z\sqrt{\pi})$.  

Our experimental results are shown in Figs.~\ref{fig:results}b-c. We also compare the theoretical shielded decay rate (solid blue) with the one we would expect in the same crêpe geometry if there was no elastic dipolar potential to repel the atoms (dashed blue). In contrast to the loss rate in a 3D geometry (red), which increases with $\sqrt{B}$ (see Supplementary Information for comments about this scaling), we observe the signature of shielding in Fig.~\ref{fig:results}b: a much weaker dependence on magnetic fields (solid blue).

We operate in the quasi-2D regime which differs from the pure-2D one in several aspects. Compared to pure-2D, the finite axial extent of the quasi-2D geometry softens the radial barrier, reducing the barrier height to energies comparable to typical temperatures in the experiment. Furthermore, for Zeeman energies that are larger than the axial trapping frequency, new collisional channels open, with a portion of the released energy converted into axial excitation and the remainder into radial motion. As a result, the relaxation for these processes is shifted to larger distances, thereby weakening the shielding. The first channel opening is visible in Fig.~\ref{fig:results}b around 100~mG as well as in Fig.~\ref{fig:plotTheory}a-b. The aforementioned factors lead to a relaxation rate that does not decrease with magnetic field, as it would in the pure-2D case, but instead shows a weaker increase compared to the case without a dipolar barrier (dashed blue). Fig.~\ref{fig:results}c shows the loss rate coefficient as a function of axial confinement. The loss rate decreases with confinement due to enhanced shielding by the dipolar repulsion and closing axially excited states channels.

We have reduced the loss rate coefficient to approximately $1 \times 10^{-12}$~cm${^3}/$s. Over a large range of magnetic fields in a lattice, we achieved more than an order of magnitude reduction in the dipolar relaxation rate coefficient compared to the unshielded case. The agreement between the numerical calculations and the experiment enables extrapolation beyond the current limitation of the experiment: very favorable loss rate coefficients of $2 \times 10^{-13}~\mathrm{cm^3/s}$ can be achieved at 200~mG with an axial confinement of 500~kHz at 1~$\mu$K. This matches the lowest rate obtained with fermions through Pauli suppression in reference~\cite{Lev15}. Under such conditions, axial excitations are energetically forbidden and the 2D decay rate is less than a factor of 3 above the pure-2D limit. By lowering the temperature to 100~nK, the relaxation rate would be suppressed by an additional factor of three and reach the $10^{-14}~\mathrm{cm^3/s}$ regime. To further understand how these numbers are computed, we describe our theoretical model in the following paragraphs.

\section*{Theoretical model}
Dipolar relaxation rates can be calculated from Fermi's golden rule. The decay rate $\Gamma$ of 2 particles is given by
\begin{equation}
\label{eq:FermiGoldenRule}
\hbar \Gamma = 2\pi\left\vert\bra{\Psi_{\text{out}}}\hat V_\rm{dd}\ket{\Psi_{\text{in}}}\right\vert^2\rho(E),
\end{equation}
where $\rho(E)$ is the final density of states at energy $E$. The incoming wave function is an excited Zeeman state with transverse momentum $\vec{k}_i$ in the lowest harmonic oscillator state, $n_i = 0$. The outgoing wave function is a lower Zeeman state with momentum $\vec{k}_f$ in the harmonic oscillator state $n_f$. The loss rate coefficient $\beta_\mathrm{2D}$ is related to $\Gamma$ through $\beta_\mathrm{2D} =\pi L^2\Gamma$, with $L$ being the radius of the transverse box used to normalize the wave functions. The atoms are coupled by the magnetic dipole-dipole interaction:
\begin{equation}
\label{eq:dipoleOperator}
    \hat V_\rm{dd} = \frac{\mu_0}{4\pi}(g_J\mu_B)^2\frac{\hat{\vec{J}}_1 \cdot \hat{\vec{J}}_2 - 3(\hat{\vec{J}}_1\cdot\vec{u}_r)(\hat{\vec{J}}_2\cdot\vec{u}_r)}{r^3},
\end{equation}
where $\vec{r}$ is the interatomic separation (with corresponding unit vector $\vec{u}_r$). The magnetic field points along $z$. Atoms in the initial spin state $\ket{j_0} = \ket{m_{J_1} = 8, m_{J_2} = 8}$ can collide and remain in the same spin state, or relax to either $\ket{j_1} = \left(\ket{7,8}+\ket{8,7}\right)/\sqrt{2}$ or $\ket{j_2} = \ket{7,7}$. The dipole-dipole operator acting on $\ket{j_0}$ is:
\begin{eqnarray}
\label{eq:dipoleOperatorOnState}
    \hat V_\rm{dd}\ket{j_0} &=& \frac{\mu_0(Jg_J\mu_B)^2}{4\pi r^3}\left[\left(1-3\bar{z}^2\right)\ket{j_0}
    - \frac{3\bar{z}\bar{r}_+}{J^{1/2}}\ket{j_1} - \frac{3\bar{r}_+^2}{2J}\ket{j_2}\right] \\
    &=& V_{\rm{dd}, 0}\ket{j_0} + V_{\rm{dd}, 1}\ket{j_1} + V_{\rm{dd},2}\ket{j_2}
\end{eqnarray}
with $\bar z = z/r$ and $\bar r_+ = (x+iy)/r$. Equation~(\ref{eq:dipoleOperatorOnState}) shows the three effects of the dipolar interaction: an elastic scattering process, a single spin-flip proportional to $\bar z$, and a double spin-flip which implicitly depends on $z$ through $r$.

In the two-dimensional limit where $z = 0$, the single spin-flip term vanishes and the elastic term is a purely repulsive $1/\rho^3$ potential (where $\rho = \sqrt{x^2+y^2}$). This potential has an analytic solution at zero temperature ($ k_i = 0$) \cite{Ticknor09}, while other cases have to be solved numerically.

We assume a quasi-2D geometry where we ignore the effect of $V_{\rm{dd}, 0}$ on the $z$ motion, which is then factorized and described by harmonic oscillator wave functions (see Methods for a discussion on this approximation). The elastic portion of the operator in equation~(\ref{eq:dipoleOperatorOnState}) is averaged over the $z$ direction. This leads to an effective repulsive potential (see Fig.~\ref{fig:plotWavefunctions_singleChannel}a) in the one-dimensional radial Schr\"odinger equation:
\begin{equation}
\label{eq:differentialEquation}
     \left\{\frac{\hbar^2}{2\mu}\left(-\frac{d^2}{d \rho^2} + \frac{m^2-1/4}{\rho^2}\right) + \bra{n} V_{\rm{dd},\, 0}\ket{n}\right\} \phi = \frac{\hbar^2 k_i^2}{2\mu} \phi.
\end{equation}
Here, the state $\ket{n}$ is the $n^{\rm{th}}$ harmonic oscillator's state along $z$. We focus on incoming states with zero projection of orbital angular momentum, $m_i=0$, as this channel dominates for any reasonable magnetic field (see Supplementary Information).

We solve the Schr\"odinger equation for the radial wave function using numerical techniques, and use it to perturbatively calculate the dipolar relaxation rate with Fermi's golden rule~(\ref{eq:FermiGoldenRule}). In Fig.~\ref{fig:plotWavefunctions_singleChannel} we show how dipolar repulsion (Fig.~\ref{fig:plotWavefunctions_singleChannel}a) modifies the incoming wave function (Fig.~\ref{fig:plotWavefunctions_singleChannel}b) and reduces the integral of the transition matrix element (Fig.~\ref{fig:plotWavefunctions_singleChannel}c).

\begin{figure}[h]
    \centering
    \includegraphics[width=1\columnwidth]{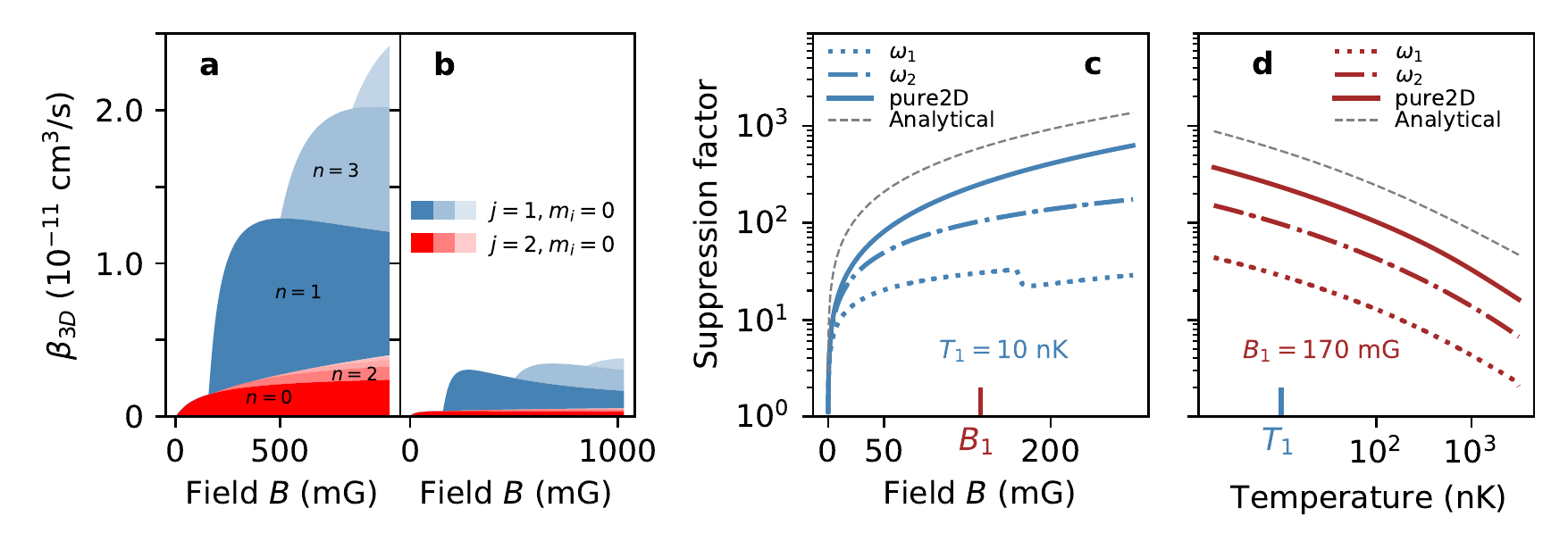}
    \caption{\textbf{Theoretical loss rate coefficients.} (a-b) Channel-by-channel decomposition of the dipolar relaxation rates of the $m_i = 0$ incoming state (valid when $\mu_B B\gg k_BT$, see Supplementary Information) in a $300~\mathrm{kHz}$ trap, both for free wave functions (a) and shielded ones (b). The blue and red colors correspond to single and double spin flips, respectively. The different shades correspond to different harmonic oscillator states as they open up with increasing magnetic field. (c-d) The suppression factor defined as the ratio of $\beta_\mathrm{2D}$ obtained from shielded and free wave functions, both at fixed temperature $T_1$ (c), and at fixed magnetic field $B_1$ (d). For each of the graphs we present curves for $\omega_1/2\pi= 300~\mathrm{kHz}$ (dotted), $\omega_2/2\pi = 1.8~\mathrm{MHz}$ (dashed dotted), the pure-2D case (solid line) as well as the analytical approximation (grey dotted) from equation~(\ref{eq:betaPure2D_Shielded}) detailed in Supplementary Information.}
    \label{fig:plotTheory}
  \end{figure}

Without axial excitation, only double spin flips to the final spin state ${\ket{j_2} = \ket{7,7}}$ and orbital state $m_f=2$ are allowed. At sufficiently high magnetic field the energy released during the collision can exceed $\hbar\omega_z$, thereby opening up new collisional channels resulting in axial excitations. Energy conservation requires
\begin{equation}
    \label{eq:energyConservation}
    \frac{\hbar^2 k_f^2}{2\mu} = \frac{\hbar^2 k_i^2}{2\mu} +\Delta j\mu_B g_J B - \Delta n \hbar \omega_z.
  \end{equation}
The single spin-flip channel ($\Delta j=1$) requires odd $\Delta n$ due to the odd symmetry of the $\bar z$ term in equation~(\ref{eq:dipoleOperatorOnState}), whereas double spin flips ($\Delta j = 2$) require even $\Delta n$. Newly opened channels increase the decay rate, as shown in Fig.~\ref{fig:plotTheory}a--b. Furthermore, as previously explained, they also decrease the shielding factor, as visible in the small notch in Fig.~\ref{fig:plotTheory}c.

Remaining in the ground state of the harmonic oscillator is therefore necessary for obtaining extremely low relaxation rates, but that requires working at low enough fields. Unfortunately, the relaxation rates we measure at very low fields deviate from the theoretical values in Fig.~\ref{fig:results}b, most likely because of imperfect circular polarization of the lattice and compensating beams. The mixture of $\sigma^+$ and $\sigma^-$ light induces Raman couplings between $\ket{m_J = +8}$ and other even $\ket{m_J}$ states, thereby opening additional relaxation channels via spin exchange \cite{Hensler03}. With a $>\!95\mathrm{\%}$ circular polarization purity, we find agreement between experimental decay rates and calculated dipolar relaxation rates for fields $>\!100~\mathrm{mG}$, where the Raman coupling is suppressed by Zeeman detuning.

\section*{Discussion and Outlook}
We have shown that confinement in thin layers not only reduces the number of available collisional channels, but additionally provides dipolar shielding, thereby strongly suppressing dipolar relaxation between atoms. In principle, arbitrarily low loss rates and infinite shielding factors are possible at very low temperatures. Strong magnetic fields are also predicted to reduce the shielded collision rate to arbitrary low values if strong axial confinement suppresses the opening of collision channels.  As we have discussed above, rather straightforward improvements in axial confinement, purity of polarization and temperature should result in rate coefficients in the $10^{-14}~\mathrm{cm^3/s}$ regime.

Our simulations and experiments show that there is already substantial shielding at thermal energies comparable to the barrier height. Lowering the temperature well below the barrier eventually results in exponential suppression~\cite{Julienne10}. For our experimental parameters, going from $1~\mathrm{\mu K}$ to $100~\mathrm{nK}$ would increase the suppression by a factor of three.

In this work, we have discussed the interplay between the elastic and inelastic aspects of dipolar interactions. Both scale with the dipolar length, which could be 10,000 times larger for polar molecules. Yet the large total angular momentum $J=8$ works in favors of dysprosium over molecules, as the elastic part of the dipole-dipole potential scales as $J^4$ in a stretched state, while the relaxation rate scales as $J^3$ for single spin-flips and $J^2$ for double spin-flips.

An important point of comparison is the elastic scattering rate. At 1 Gauss in a trap with a 2~MHz axial frequency, the inelastic 2D cross-section would be 20~nm without shielding.  Shielding drops this number to 0.3~nm, while the semi-classical dipolar elastic collisional cross section is $\sigma_{\text{SC}} = 180$~nm \cite{Ticknor09}. Shielding is necessary to obtain a ratio of good to bad collisions in excess of 100.

Dipolar shielding has previously been observed in polar molecules with fermionic statistics~\cite{valtolina2020dipolar}, for which the shielding is qualitatively different. Since identical fermions already have an isotropic $p$-wave barrier, adding moderate dipolar interactions in a confined geometry will first strengthen this barrier in the radial direction but also weaken it in the axial one. As a result, the inelastic collision rate will first decrease with the dipole moment and then increase~\cite{quemener2011dynamics}. This cannot be seen with bosons. For both particle types, the inelastic collision rate will eventually decrease when entering more deeply into the 2D regime, as we have explored in this work. Our technique would be crucial to study spin mixtures of bulk gases of bosonic dipolar species.

In conclusion, we have demonstrated a way to realize long-lived spin mixtures in dense bosonic lanthanide clouds, opening up new possibilities for quantum simulation experiments in two dimensions. With such technique, dysprosium can be used to study quantum materials with dipolar interactions in regimes different from those currently possible for polar molecules \cite{Zoller07} and Rydberg atoms \cite{Browaeys20}. Stable spin mixtures are important for implementing spin-orbit coupling and artificial gauge potentials via Raman coupling of spin states \cite{Lin09, Lin11}. By suppressing dipolar relaxation, one can take advantage of the ground state orbital angular momentum of lanthanides to avoid the substantial photon scattering rates of the Raman beams for alkali atoms \cite{burdick2016long}.

\bmhead{Acknowledgments} We thank Brice Bakkali-Hassani, Hanzhen Lin and Yu-Kun Lu for comments on the manuscript. We acknowledge support from the NSF through the Center for Ultracold Atoms and through Grant No. 1506369, the Vannevar-Bush Faculty Fellowship, and an ARO DURIP grant.

\bmhead{Author contributions} P.B., M.C, L.D., W.L., A.O.J. and W.K. designed and constructed the experimental setup, P.B., M.C., L.D. and J.d.H carried out the experimental work, P.B., M.C., L.D., J.d.H and W.K. developed the theoretical models and simulations, all authors contributed to the writing of the manuscript.

\bmhead{Competing interests} The authors declare no competing interests.

\clearpage

\setcounter{equation}{0}
\setcounter{table}{0}
\makeatletter
\renewcommand{\theequation}{S\arabic{equation}}

\setcounter{figure}{0}
\makeatletter
\renewcommand{\thefigure}{S\arabic{figure}}
\renewcommand{\bibnumfmt}[1]{[S#1]}
\renewcommand{\citenumfont}[1]{S#1}

\section*{Methods}

\subsection*{Sample preparation}
We prepare spin-polarized samples of $\sim\!8\times 10^4$ $^{162}\rm{Dy}$ atoms in the $\ket{J = 8, m_J = -8}$ state in an optical dipole trap just above the transition temperature. The samples are obtained after evaporative cooling in a crossed optical-dipole trap (ODT) which is loaded from the narrow-line magneto-optical trap described in reference~\cite{Lunden20}. Working with a thermal gas makes it easier to determine dipolar relaxation rate coefficients without accounting for a varying condensate fraction.

The highest spin state $\ket{m_J = +8}$ is populated via adiabatic rapid passage using an RF sweep in a magnetic field of $3.5~\rm{G}$ along the $z$ direction. A stack of quasi-2D layers, which we refer to as crêpes due to their extreme aspect ratio, is created using a 1D optical lattice formed by retroreflecting a 741~nm laser beam along the $z$ axis. The lattice beam is blue-detuned from the 741~nm transition (which has a linewidth $\Gamma/2\pi = 1.8~\mathrm{kHz}$) by several GHz, thus providing frequency-controllable tight axial confinement. A coaxial vertical optical dipole trap is used to compensate for the transverse repulsion resulting from the blue-detuned lattice. The lattice and the vertical dipole trap are turned on using exponential ramps with a $50~\rm{ms}$ time constant to adiabatically load the atoms into the lowest vibrational level of the 2D layers. During the first $40~\rm{ms}$ of the lattice ramp, the magnetic field is rapidly reduced to $40~\rm{mG}$ to minimize the dipolar relaxation losses. The magnetic field is then ramped up to its final value during the last $10~\rm{ms}$ of the lattice loading ramp, after which the decay of the sample due to inelastic collisions is measured.

\subsection*{Zeroing the magnetic field}
\label{appendix:zeroMagneticField}
\begin{figure}[h]
  \centering
  \includegraphics[width=0.7\columnwidth]{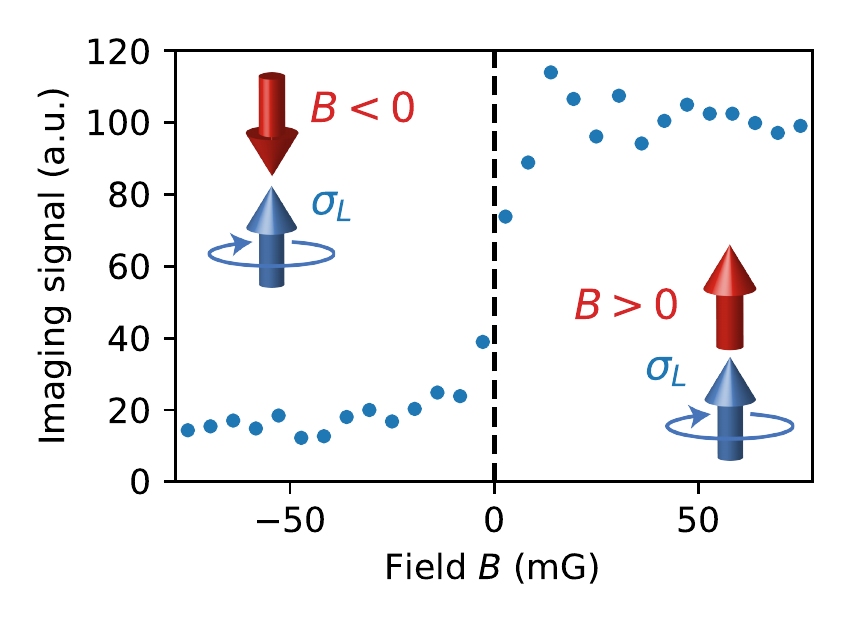}
  \caption{\textbf{Determining the zero of the magnetic field.}  For spin-polarized $m_J = -8$ dysprosium atoms and left-circularly polarized imaging light, the  drastic difference in Clebsch-Gordan coefficients for $\sigma_+$ and $\sigma_-$ transitions produces a step-like change in imaging signal as the magnetic field traverses through zero.}
  \label{fig:magnetic_zeroing}
\end{figure}

Achieving control of low magnetic fields is critical for minimizing dipolar suppression by preventing higher outgoing vibrational channels from opening.  We have devised a method to zero the magnetic field that relies on the large disparity of Clebsch-Gordan coefficients for dysprosium.  When an atom's magnetic moment is aligned along the propagation of a circularly polarized imaging beam, the amount of scattered light strongly differs whether the  magnetic dipole moment is oriented parallel or  anti-parallel to the propagation of the imaging beam. By using absorption imaging for various external magnetic fields, as shown in Fig.~\ref{fig:magnetic_zeroing}, one can observe when the dipole moment has flipped, which determines the zero of the external magnetic field.

More specifically, in a spin-polarized ($m_J = -8$) sample of bosonic dysprosium, the Clebsch-Gordan coefficients for $\sigma_-$, $\pi$ and $\sigma_+$ transitions are 1, 1/9 and 1/153 respectively. We perform absorption imaging of a spin-polarized sample with left-circularly polarized ($\sigma_L$) light along the magnetic field quantization axis $z$. We work with low enough light intensity and imaging time to prevent optical pumping. At large positive magnetic field bias, the atoms see $\sigma_-$ light with a corresponding Clebsch-Gordan coefficient of 1, resulting in a large atom count. At large negative magnetic field bias, the atoms see $\sigma_+$ light with a corresponding Clebsch-Gordan coefficient of 1/153 leading to a low atom count. The lower the transverse magnetic field, the sharper is the transition when the longitudinal field is varied.  In this way, the zero settings for all components of the magnetic field are determined.

\subsection*{Lattice light choice}
The need for deep optical lattices requires a tightly focused lattice beam, which causes undesirably strong radial confinement if one uses a red-detuned beam. By choosing a blue-detuned lattice we avoid adiabatic compression of the cloud in the transverse direction and the substantial corresponding increase in temperature when ramping up the optical lattice. The choice of a blue-detuned lattice also exposes the atoms to lower light intensities and reduces the unwanted Raman transitions due to imperfect circular polarization. However, the radial deconfinement created by the lattice needs to be compensated, which we achieve with a red-detuned optical dipole trap that enables independent control of the axial and transverse trap frequencies (see Fig.~\ref{fig:plotLatticeGeometry} left).

The lattice was created by near-resonant 741~nm light from a Ti:Sapph laser which can deliver about 300~mW of light to the atoms, after fiber coupling and intensity stabilization.

\subsection*{Trap geometry}
\label{appendix:latticeGeometry}
Atoms are loaded into an optical dipole trap consisting of three $1064~\rm{nm}$ laser beams: two beams with $40~\rm{\mu m}$ beam waists crossed at $8^\circ$ in the horizontal plane, and a beam with a $64~\rm{\mu m}$ waist propagating along the (vertical) $z$ direction. During the dipolar relaxation experiment, the horizontal beams are switched off, and the vertical beam serves to compensate for the deconfinement of the blue-detuned lattice. The lattice beam is focused down to a waist of 50\,$\mu$m. It is typically detuned by 14.25 to 2.25\,GHz to the blue side of the narrow 1.8\,kHz transition \cite{lu2011spectroscopy}. The transverse antitrapping potential is compensated using 8\,W in the vertical trapping beam. We verified with in-situ images (obtained with detuned imaging light due to the high optical densities) that the blue-detuned lattice is correctly compensated without displacement of the cloud.

\begin{figure}[h]
    \centering
    \includegraphics[width=0.7\columnwidth]{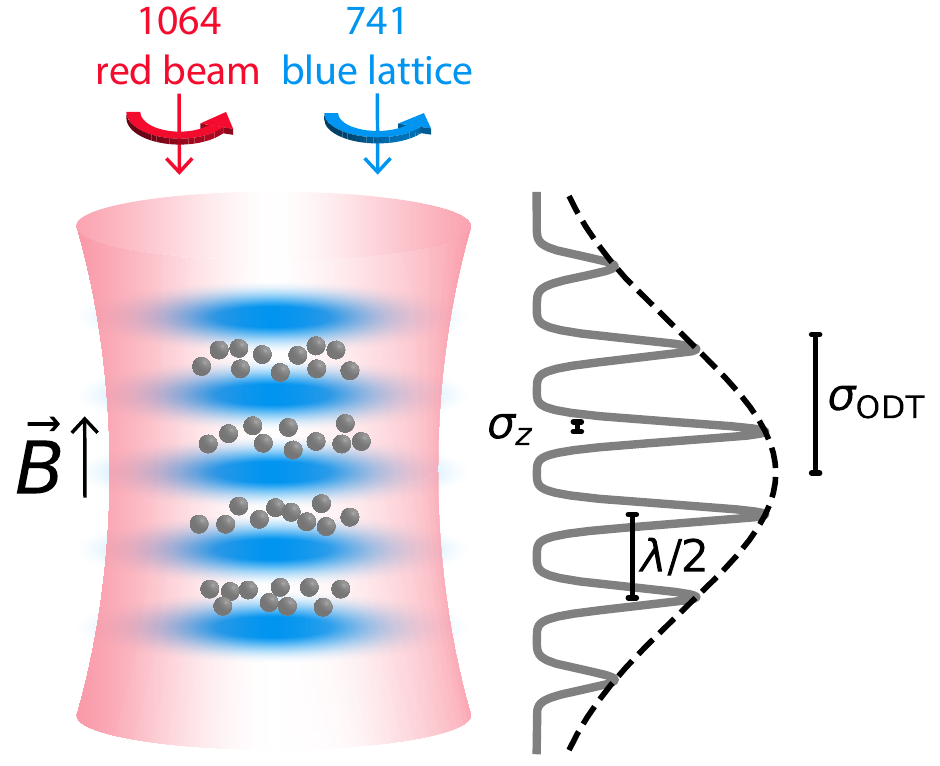}
    \caption{\textbf{Trap geometry and  relevant length scales.} Left: Reproduction of Fig.~\ref{fig:results}a of the main text. Right: The spatial density of the cloud in the longitudinal direction is characterized by the axial RMS width $\sigma_z = a_z/2$, the lattice spacing $\lambda/2$ and the initial width of the loaded thermal cloud $\sigma_\rm{ODT}$.}  
    \label{fig:plotLatticeGeometry}
  \end{figure}

The RMS extension of the cloud along the lattice direction before loading is $\sigma_\mathrm{ODT}\simeq 4.7\, \mu$m. Given the layer separation of $\lambda/2 \simeq 371$\,nm, around $4\sqrt{\pi}\sigma_\rm{ODT}/\lambda = 45$ crêpes are loaded with initially $3\times 10^4$ atoms and a central density of $n_0 = 2.9\times 10^9\, \text{cm}^{-2}$. The density distribution in the $i^{\text{th}}$ pancake is described by (see Fig.~\ref{fig:plotLatticeGeometry})
\begin{equation}
    n_i(t, \rho, z) = n_0(t) \exp\left(-z_i^2/(2\sigma_{\rm{ODT}}^2) \right)\exp\left(-\rho^2/(2\sigma_{\perp}^2) \right)
    \label{eq:density}
\end{equation}
with $z_i = i\frac{\lambda}{2}$ and $\sigma_{\perp} = \sqrt{\frac{k_B T_\mathrm{lattice}}{2\mu\omega_\perp^2}}$, $\sigma_\mathrm{ODT} = \sqrt{\frac{k_B T_\mathrm{ODT}}{2\mu\omega_\mathrm{ODT}^2}}$. The parameters $\omega_\mathrm{ODT} = 2\pi\cdot94$\,Hz and $T_\mathrm{ODT}= 150$\,nK describe the cloud before the lattice is ramped up whereas $\omega_\perp = 2\pi\cdot200$\,Hz and $T_\mathrm{lattice}\simeq 1\,\mu$K characterize the conditions after lattice ramp up.  The central crêpe contains about 900 atoms. The RMS width of the crêpes is typically $\sigma_z \simeq 10 $\,nm while the radial one is $\sigma_\perp = 5.7\,\mu$m.

\subsection*{Lifetime analysis}
\label{appendix:lifetimeAnalysis}

The decay of the cloud can be described via equation~(\ref{eq:two-body-decay-diff_eq}) for the 3D densities
\begin{equation}
    \frac{\d n_\mathrm{3D}}{\d t} = -\beta_\mathrm{3D}n_\mathrm{3D}^2.
    \label{eq:differential_eq_3D}
\end{equation}
or by using a 2D equation
\begin{equation}
    \frac{\d n_\mathrm{2D}}{\d t} = -\beta_\mathrm{2D} n_\mathrm{2D}^2
    \label{eq:differential_eq_2D}.
\end{equation}
The densities in each pancake are related by
\begin{equation}
    n_\mathrm{3D} = n_\mathrm{2D}\frac{1}{\sqrt{2\pi}\sigma_z}\exp\left(-z^2/(2\sigma_z^2)\right)
\end{equation}
with $\sigma_z = \sqrt{\frac{\hbar}{4\mu\omega_z}} = a_z/2 \simeq 10~\text{nm}$. When integrating equation (\ref{eq:differential_eq_3D}) and equating it to (\ref{eq:differential_eq_2D}), we obtain
\begin{equation}
    \beta_\mathrm{3D} =  2\sqrt{\pi}\sigma_z\beta_\mathrm{2D}
    \label{eq:beta2Dto3D}.
\end{equation}
In the main paper, we are using $\beta_\rm{3D}$ to characterize the decay.

We will omit the 2D subscript for the densities in the rest of the manuscript. Equation~(\ref{eq:differential_eq_2D}) -- in a local-density approximation -- needs to be integrated over the cloud volume to relate to the observed quantity $N$, the number of atoms:
\begin{equation}
\label{eq:differentialEquation_decay}
    \frac{dN}{dt} = -\beta_\mathrm{2D}\int_{\mathcal{S}}n^2\mathrm{d}\tau = -\beta_\mathrm{2D}\left<n\right> \equiv -\beta_\mathrm{2D}\frac{N^2}{V_\mathrm{eff}}.
\end{equation}

The effective volume $V_\rm{eff}$ is determined as follows. After integration of equation~(\ref{eq:density}), one gets $N = \sum_i N_i = \sum_i \int n_i\d \tau = n_0 2\pi\sigma_\perp^2\sqrt{2\pi}\sigma_{\mathrm{ODT}}/(\lambda/2)$ and
\begin{multline*}
\frac{\d N}{\d t} = \sum_i\frac{\d N_i}{\d t} = -\beta_\mathrm{2D} \sum_i \int_\mathcal{S}n_i^2 \: 2\pi\rho\d\rho = \: -\beta_\mathrm{2D}n_0^2\pi\sigma_\perp^2\sum_i e^{-\frac{z_i^2}{2\sigma_\mathrm{ODT}^2}}  \\ = -\beta_\mathrm{2D}n_0^2\pi\sigma_\perp^2 \frac{\sqrt{\pi}\sigma_{\mathrm{ODT}}}{\lambda/2}.
\end{multline*}
Identifying $V_\mathrm{eff}$ in equation~(\ref{eq:differentialEquation_decay}) gives
\begin{equation*}
    V_\mathrm{eff} = 4\pi\sigma_\perp^2\frac{2\sqrt{\pi}\sigma_\mathrm{ODT}}{\lambda/2}.
\end{equation*}

We note that $\left<n\right> = \frac{N}{2^{3/2}V_\mathrm{eff}}$, where each $\sqrt{2}$ factor comes from the Gaussian averaging along one axis. To take into account the moderate heating during the experiment, we perform a linear fit of the temperature $T(t)=T_0+v_Tt$ which is used to scale the effective volume $V_\rm{eff}(t)$. The solution of the differential equation~(\ref{eq:differentialEquation_decay}) that we fit is $N(t) = \frac{N_0}{1+\frac{\beta_\mathrm{2D}}{V_\mathrm{eff}}N_0 \frac{\ln\left(1+v_T t\right)}{v_T}}$ from which we determine $\beta_\mathrm{2D}$. The atom number $N(t)$ is measured as a function of hold time (typically tens of ms) using time of flight imaging.

Here we have assumed that every dipolar relaxation event leads to the loss of both atoms. This is justified since the effective trap depth of a few micro-kelvins is negligible compared to the kinetic energy gained by the spin-flip for magnetic fields larger than a few tens of milligauss. The experiment is sufficiently fast (tens of ms) such that photon scattering in the lattice, background collisions and residual evaporation are not important. 

\subsection*{Error bars}

The uncertainties represented by the errorbars in the plots come from the statistical error due to the curve fitting, as well as our best estimate in the uncertainties of $\omega_\mathrm{ODT}$, $T_\mathrm{ODT}$, $\omega_\perp$,  $\omega_z$ and $T_\mathrm{lattice}$. $\omega_\mathrm{ODT}$ is measured in the ODT by observing the oscillation of the cloud after suddenly applying a magnetic force.  $\omega_\perp$ is measured by modulating the intensity of the vertical trapping beam and observing the parametric heating resonance. Axial trap frequencies $\omega_z$ (which go up to several hundreds of kHz) are also measured via parametric heating.  Due to the bandwidth of the drive electronics, this was done only in shallow lattices and extrapolated to deeper lattices.  Temperatures are observed in time of flight.

\subsection*{Theoretical decay rate}
We summarize here the main equations to produce the theoretical predictions in Figs.~\ref{fig:plotWavefunctions_singleChannel}, \ref{fig:results} and \ref{fig:plotTheory}. Detailed derivations are given in Supplementary Information.

We recall equation (\ref{eq:dipoleOperatorOnState})
\begin{equation}
\label{eq:dipolePotentialMethods}
    \hat V_\rm{dd}\ket{j_0} = \frac{\mu_0(Jg_J\mu_B)^2}{4\pi r^3}\left[\left(1-3\bar{z}^2\right)\ket{j_0}
    - \frac{3\bar{z}\bar{r}_+}{J^{1/2}}\ket{j_1} - \frac{3\bar{r}_+^2}{2J}\ket{j_2}\right]
\end{equation}
which is used to compute the potential used in equation (\ref{eq:differentialEquation}):
\begin{equation}
     \left\{\frac{\hbar^2}{2\mu}\left(-\frac{d^2}{d \rho^2} + \frac{m^2-1/4}{\rho^2}\right) + \mel{n}{V_{\rm{dd},\, 0}}{n}\right\} \phi = \frac{\hbar^2 k_i^2}{2\mu} \phi.
\end{equation}
This is the equation that we solve numerically for both the incoming ${(m=0\text{, }n=0)}$ and outgoing ($m=2$, $n$ even or $m=1$, $n$ odd) wave functions. The code we developed combines grids of multiple step-sizes to account for the need to appropriately average the potential along $z$, describe the short-range shielding at small $\rho$ and normalize correctly the wave functions at large distances. Given the temperature, magnetic fields, $z$ trapping frequencies and desired precision, the code determines an appropriate grid, and computes the incoming wave function and the harmonic oscillator states on this specific grid. It then distributes those results on multiple cores, computing the outgoing wave function for each of the different decay channels and the respective integral of Fermi's golden rule. This method enables the code to produce the plots presented in this paper on a simple laptop in a reasonable time.

The normalization condition reads: $\int_{0}^{L} \d \rho\, \phi_{n,m}(\rho)^2 = 1$ for a cylinder of radius $L$. Our model accounts for the modification of both incoming and outgoing wave functions by the dipolar interaction. The free radial wave function solution with momentum $k$ is $\phi_{n, m}^{(\mathrm{free})}(\rho) = \sqrt{\frac{\pi k \rho}{L}}J_{m}(k \rho)$, which does not depend on $n$.

The 2D loss rate coefficient for the channel ${\ket{j_0}\rightarrow \ket{j_f}, \ket{0} \rightarrow \ket{n_f}}$ reads
\begin{equation*}
\beta_\rm{2D}^{j_f, n_f} = \frac{8\mu}{k_i k_f\hbar^3}\left\vert L\int_{-\infty}^{+\infty}\d z\int_{0}^{L} \d\rho  \phi_{n_f, j_f}(\rho)\chi_{n_f}(z)V_{\rm{dd},\, j_f}(\rho, z)\chi_{0}(z)\phi_{0}(\rho)\right\vert^2
\end{equation*}

with $\chi_n$ being the $n^{\rm{th}}$ harmonic oscillator's state wave function:
\begin{equation*}
\chi_n(z) = \frac{1}{\sqrt{2^n n!}}\left(\frac{1}{\pi a_z^2}\right)^{1/4}H_n(z/a_z)e^{-\frac{z^2}{2a_z^2}}.
\end{equation*}

The total rate is then the sum over all channels:
\begin{equation*}
\beta_{\rm{2D}} = \sum_{j_f, n_f}\beta_\rm{2D}^{j_f, n_f}.
\end{equation*}
and relates to the 3D rate as $\beta_\rm{3D} = \sqrt{\pi}a_z\beta_\rm{2D}$.

The rate is eventually averaged over the thermal distribution of incoming momenta (see Supplementary Information) for the Fig.~\ref{fig:results}~b-c, and computed at the mean momentum for all of the other figures.

\subsection*{Pure-2D limit}
In pure-2D the double spin-flip potential is $V_{\rm{dd}, 2}(\tilde{\rho}) \propto 1/\rho^3$.
If we ignore the shielding, in the low-temperature limit, we find that the 2D decay rate is
\begin{equation}
    \beta^\text{pure-2D}_\rm{free} = 4\pi^2\frac{1}{J^2}\frac{E_\rm{dd}}{\hbar}a_\rm{dd}^4 k_f^2.
\end{equation}
So $\beta^\text{pure-2D}_\rm{free}\propto B$. If we incorporate shielding we find that
\begin{equation}
\beta^\text{pure-2D}_\mathrm{shielded} \propto \left(1/\log(k_i)\right)^2
\end{equation}
which goes to zero at zero temperature. Under certains assumptions detailed in Supplementary Information and noting $x_2$ the first zero of the Bessel function $J_2(x)$, one can find that the decay rate in a certain field range behaves as
\begin{equation}
    \beta^\text{pure-2D}_ \mathrm{shielded} \propto k_f^{1/4}\exp\left(-2\sqrt{\frac{8a_\rm{dd} k_f}{x_2}}\right),
\end{equation}
which vanishes at high magnetic fields.

\subsection*{Discussion of various approximations}

\textbf{Unitarity limit}. The perturbative results will get modified when the decay rate approaches the unitary limit.  However, in our range of parameters, the decay rates are much smaller than the unitary limit. A $\beta_\rm{3D}$ of high $10^{-12}\rm{cm}^3/$s corresponds to a $\beta_\rm{2D}=\sigma\hbar k/\mu$ in the low $10^{-6}\rm{cm}^2/$s. This gives a $\sigma k\simeq 0.1 \ll 4$ which puts us safely in the non-unitary regime \footnote{The total cross section in 2D is $\sigma = \frac{4}{k}\sum_m\sin^2\delta$, and we are dominated by the $s$-wave contribution given our magnetic fields.}.

\textbf{Wave function substitution approximation}. The system is perturbed by two part of the dipolar potential: one which is diagonal in the spin states basis and therefore elastic, the other part is non-diagonal and causes transitions. Usually, the weakest part of the Hamiltonian should be treated perturbatively. It is the case here since $\left\vert\hat V_\rm{dd}^\rm{inelastic} / \hat V_\rm{dd}^\rm{elastic}\right\vert^2\simeq 1/J = 1/8$ as shown in equation~(\ref{eq:dipolePotentialMethods}). This is why we evaluate the decay rate on the shielded wave function. Following previous treatments \cite{Hensler03, Pasquiou10, Lev15}, we have not checked the importance of higher-order terms in the perturbation theory.

\textbf{Effective potential approximation}. To compute the wave functions we assumed an effective potential obtained by averaging $V_\rm{dd}$ in the $n^{\text{th}}$ state of the harmonic oscillator. This is the diabatic limit of a coupled channels calculation. There exists a fully adiabatic method to compute the molecular potential of two interacting dipoles in a quasi-2D geometry~\cite{Ticknor10}. It would mix the harmonic oscillator states but we found it would only affect the wave function at short distances, which is important only at high magnetic fields. In our experiment, $\tilde k_f$ remains on the order of 1, and restricting the $z$-motion to the pre-existing harmonic oscillator states is acceptable.

\textbf{Fermi's golden rule approximation}. The use of Fermi's golden rule with the original density distribution is valid only if the decay rate is smaller than the other time constants of the system. The relaxation rate $\Gamma = \beta_\mathrm{3D}n\simeq 10^2$\,s$^{-1}$ is indeed smaller than the collision rate which is around $10^3$\,s$^{-1}$, or the trap frequencies of 200~Hz. This assumption is therefore fulfilled. The system will stay in (quasi-) equilibrium when the loss rates are smaller than the trapping frequencies and smaller than the rate of elastic collisions which provides thermalization. 

\textbf{Neglecting molecular potentials}. The background $s$-wave scattering length of dysprosium $a = 5.9$~nm can modify the wave functions. A previous paper \cite{Pasquiou10} studied extensively its influence on chromium. However, the decay rate we observed in a large volume 3D trap (red curve in Fig.~\ref{fig:results}) agrees better with the theory which does not take the scattering length into account. Another dysprosium experiment \cite{Lev15} found a similar result in an even wider range of fields. However, it is possible that the short-range molecular potential plays a role in the 2D results, and could possibly explain why we obtain rates a few times smaller than the theory predicts (see shaded areas in Fig.~\ref{fig:results}). Indeed, a sizeable contribution to the loss comes from interatomic distances smaller than the van der Waals length $a_{\text{vdW}} = 4.3$~nm and the scattering length $a = 5.9$~nm. Since the real wave function rapidly oscillates at short-range, the contribution to the overlap matrix element should vanish.

\newpage

\bibliography{dipolarRelaxation}


\begin{thebibliography}{32}
\ifx \bisbn   \undefined \def \bisbn  #1{ISBN #1}\fi
\ifx \binits  \undefined \def \binits#1{#1}\fi
\ifx \bauthor  \undefined \def \bauthor#1{#1}\fi
\ifx \batitle  \undefined \def \batitle#1{#1}\fi
\ifx \bjtitle  \undefined \def \bjtitle#1{#1}\fi
\ifx \bvolume  \undefined \def \bvolume#1{\textbf{#1}}\fi
\ifx \byear  \undefined \def \byear#1{#1}\fi
\ifx \bissue  \undefined \def \bissue#1{#1}\fi
\ifx \bfpage  \undefined \def \bfpage#1{#1}\fi
\ifx \blpage  \undefined \def \blpage #1{#1}\fi
\ifx \burl  \undefined \def \burl#1{\textsf{#1}}\fi
\ifx \doiurl  \undefined \def \doiurl#1{\url{https://doi.org/#1}}\fi
\ifx \betal  \undefined \def \betal{\textit{et al.}}\fi
\ifx \binstitute  \undefined \def \binstitute#1{#1}\fi
\ifx \binstitutionaled  \undefined \def \binstitutionaled#1{#1}\fi
\ifx \bctitle  \undefined \def \bctitle#1{#1}\fi
\ifx \beditor  \undefined \def \beditor#1{#1}\fi
\ifx \bpublisher  \undefined \def \bpublisher#1{#1}\fi
\ifx \bbtitle  \undefined \def \bbtitle#1{#1}\fi
\ifx \bedition  \undefined \def \bedition#1{#1}\fi
\ifx \bseriesno  \undefined \def \bseriesno#1{#1}\fi
\ifx \blocation  \undefined \def \blocation#1{#1}\fi
\ifx \bsertitle  \undefined \def \bsertitle#1{#1}\fi
\ifx \bsnm \undefined \def \bsnm#1{#1}\fi
\ifx \bsuffix \undefined \def \bsuffix#1{#1}\fi
\ifx \bparticle \undefined \def \bparticle#1{#1}\fi
\ifx \barticle \undefined \def \barticle#1{#1}\fi
\bibcommenthead
\ifx \bconfdate \undefined \def \bconfdate #1{#1}\fi
\ifx \botherref \undefined \def \botherref #1{#1}\fi
\ifx \url \undefined \def \url#1{\textsf{#1}}\fi
\ifx \bchapter \undefined \def \bchapter#1{#1}\fi
\ifx \bbook \undefined \def \bbook#1{#1}\fi
\ifx \bcomment \undefined \def \bcomment#1{#1}\fi
\ifx \oauthor \undefined \def \oauthor#1{#1}\fi
\ifx \citeauthoryear \undefined \def \citeauthoryear#1{#1}\fi
\ifx \endbibitem  \undefined \def \endbibitem {}\fi
\ifx \bconflocation  \undefined \def \bconflocation#1{#1}\fi
\ifx \arxivurl  \undefined \def \arxivurl#1{\textsf{#1}}\fi
\csname PreBibitemsHook\endcsname

\bibitem{wilpers2002optical}
\begin{barticle}
\bauthor{\bsnm{Wilpers}, \binits{G.}},
\bauthor{\bsnm{Binnewies}, \binits{T.}},
\bauthor{\bsnm{Degenhardt}, \binits{C.}},
\bauthor{\bsnm{Sterr}, \binits{U.}},
\bauthor{\bsnm{Helmcke}, \binits{J.}},
\bauthor{\bsnm{Riehle}, \binits{F.}}:
\batitle{Optical clock with ultracold neutral atoms}.
\bjtitle{Physical review letters}
\bvolume{89}(\bissue{23}),
\bfpage{230801}
(\byear{2002})
\end{barticle}
\endbibitem

\bibitem{Pasquiou10}
\begin{barticle}
\bauthor{\bsnm{Pasquiou}, \binits{B.}},
\bauthor{\bsnm{Bismut}, \binits{G.}},
\bauthor{\bsnm{Beaufils}, \binits{Q.}},
\bauthor{\bsnm{Crubellier}, \binits{A.}},
\bauthor{\bsnm{Mar\'echal}, \binits{E.}},
\bauthor{\bsnm{Pedri}, \binits{P.}},
\bauthor{\bsnm{Vernac}, \binits{L.}},
\bauthor{\bsnm{Gorceix}, \binits{O.}},
\bauthor{\bsnm{Laburthe-Tolra}, \binits{B.}}:
\batitle{Control of dipolar relaxation in external fields}.
\bjtitle{Phys. Rev. A}
\bvolume{81},
\bfpage{042716}
(\byear{2010})
\end{barticle}
\endbibitem

\bibitem{Cornish2000}
\begin{barticle}
\bauthor{\bsnm{Cornish}, \binits{S.L.}},
\bauthor{\bsnm{Claussen}, \binits{N.R.}},
\bauthor{\bsnm{Roberts}, \binits{J.L.}},
\bauthor{\bsnm{Cornell}, \binits{E.A.}},
\bauthor{\bsnm{Wieman}, \binits{C.E.}}:
\batitle{Stable ${}^{85}\mathrm{Rb}$ bose-einstein condensates with widely
  tunable interactions}.
\bjtitle{Phys. Rev. Lett.}
\bvolume{85},
\bfpage{1795}--\blpage{1798}
(\byear{2000})
\end{barticle}
\endbibitem

\bibitem{valtolina2020dipolar}
\begin{barticle}
\bauthor{\bsnm{Valtolina}, \binits{G.}},
\bauthor{\bsnm{Matsuda}, \binits{K.}},
\bauthor{\bsnm{Tobias}, \binits{W.G.}},
\bauthor{\bsnm{Li}, \binits{J.-R.}},
\bauthor{\bsnm{De~Marco}, \binits{L.}},
\bauthor{\bsnm{Ye}, \binits{J.}}:
\batitle{Dipolar evaporation of reactive molecules to below the fermi
  temperature}.
\bjtitle{Nature}
\bvolume{588}(\bissue{7837}),
\bfpage{239}--\blpage{243}
(\byear{2020})
\end{barticle}
\endbibitem

\bibitem{Anderegg2021}
\begin{barticle}
\bauthor{\bsnm{Anderegg}, \binits{L.}},
\bauthor{\bsnm{Burchesky}, \binits{S.}},
\bauthor{\bsnm{Bao}, \binits{Y.}},
\bauthor{\bsnm{Yu}, \binits{S.S.}},
\bauthor{\bsnm{Karman}, \binits{T.}},
\bauthor{\bsnm{Chae}, \binits{E.}},
\bauthor{\bsnm{Ni}, \binits{K.-K.}},
\bauthor{\bsnm{Ketterle}, \binits{W.}},
\bauthor{\bsnm{Doyle}, \binits{J.M.}}:
\batitle{Observation of microwave shielding of ultracold molecules}.
\bjtitle{Science}
\bvolume{373}(\bissue{6556}),
\bfpage{779}--\blpage{782}
(\byear{2021})
\end{barticle}
\endbibitem

\bibitem{chomaz2022dipolar}
\begin{botherref}
\oauthor{\bsnm{Chomaz}, \binits{L.}},
\oauthor{\bsnm{Ferrier-Barbut}, \binits{I.}},
\oauthor{\bsnm{Ferlaino}, \binits{F.}},
\oauthor{\bsnm{Laburthe-Tolra}, \binits{B.}},
\oauthor{\bsnm{Lev}, \binits{B.L.}},
\oauthor{\bsnm{Pfau}, \binits{T.}}:
Dipolar physics: A review of experiments with magnetic quantum gases.
Reports on Progress in Physics
(2022)
\end{botherref}
\endbibitem

\bibitem{lahaye2009physics}
\begin{barticle}
\bauthor{\bsnm{Lahaye}, \binits{T.}},
\bauthor{\bsnm{Menotti}, \binits{C.}},
\bauthor{\bsnm{Santos}, \binits{L.}},
\bauthor{\bsnm{Lewenstein}, \binits{M.}},
\bauthor{\bsnm{Pfau}, \binits{T.}}:
\batitle{The physics of dipolar bosonic quantum gases}.
\bjtitle{Reports on Progress in Physics}
\bvolume{72}(\bissue{12}),
\bfpage{126401}
(\byear{2009})
\end{barticle}
\endbibitem

\bibitem{Lian2012}
\begin{barticle}
\bauthor{\bsnm{Lian}, \binits{B.}},
\bauthor{\bsnm{Ho}, \binits{T.-L.}},
\bauthor{\bsnm{Zhai}, \binits{H.}}:
\batitle{Searching for non-abelian phases in the bose-einstein condensate of
  dysprosium}.
\bjtitle{Phys. Rev. A}
\bvolume{85},
\bfpage{051606}
(\byear{2012})
\end{barticle}
\endbibitem

\bibitem{Deng2012}
\begin{barticle}
\bauthor{\bsnm{Deng}, \binits{Y.}},
\bauthor{\bsnm{Cheng}, \binits{J.}},
\bauthor{\bsnm{Jing}, \binits{H.}},
\bauthor{\bsnm{Sun}, \binits{C.-P.}},
\bauthor{\bsnm{Yi}, \binits{S.}}:
\batitle{Spin-orbit-coupled dipolar bose-einstein condensates}.
\bjtitle{Phys. Rev. Lett.}
\bvolume{108},
\bfpage{125301}
(\byear{2012})
\end{barticle}
\endbibitem

\bibitem{Lian2014}
\begin{barticle}
\bauthor{\bsnm{Lian}, \binits{B.}},
\bauthor{\bsnm{Zhang}, \binits{S.}}:
\batitle{Singlet mott state simulating the bosonic laughlin wave function}.
\bjtitle{Phys. Rev. B}
\bvolume{89},
\bfpage{041110}
(\byear{2014})
\end{barticle}
\endbibitem

\bibitem{Cui2013}
\begin{barticle}
\bauthor{\bsnm{Cui}, \binits{X.}},
\bauthor{\bsnm{Lian}, \binits{B.}},
\bauthor{\bsnm{Ho}, \binits{T.-L.}},
\bauthor{\bsnm{Lev}, \binits{B.L.}},
\bauthor{\bsnm{Zhai}, \binits{H.}}:
\batitle{Synthetic gauge field with highly magnetic lanthanide atoms}.
\bjtitle{Phys. Rev. A}
\bvolume{88},
\bfpage{011601}
(\byear{2013})
\end{barticle}
\endbibitem

\bibitem{Babik20}
\begin{barticle}
\bauthor{\bsnm{Babik}, \binits{D.}},
\bauthor{\bsnm{Roell}, \binits{R.}},
\bauthor{\bsnm{Helten}, \binits{D.}},
\bauthor{\bsnm{Fleischhauer}, \binits{M.}},
\bauthor{\bsnm{Weitz}, \binits{M.}}:
\batitle{Synthetic magnetic fields for cold erbium atoms}.
\bjtitle{Physical Review A}
\bvolume{101}(\bissue{5}),
\bfpage{053603}
(\byear{2020})
\end{barticle}
\endbibitem

\bibitem{Gopalakrishnan2013}
\begin{barticle}
\bauthor{\bsnm{Gopalakrishnan}, \binits{S.}},
\bauthor{\bsnm{Martin}, \binits{I.}},
\bauthor{\bsnm{Demler}, \binits{E.A.}}:
\batitle{Quantum quasicrystals of spin-orbit-coupled dipolar bosons}.
\bjtitle{Phys. Rev. Lett.}
\bvolume{111},
\bfpage{185304}
(\byear{2013})
\end{barticle}
\endbibitem

\bibitem{Yi2006}
\begin{barticle}
\bauthor{\bsnm{Yi}, \binits{S.}},
\bauthor{\bsnm{Pu}, \binits{H.}}:
\batitle{Spontaneous spin textures in dipolar spinor condensates}.
\bjtitle{Phys. Rev. Lett.}
\bvolume{97},
\bfpage{020401}
(\byear{2006})
\end{barticle}
\endbibitem

\bibitem{Lev15}
\begin{barticle}
\bauthor{\bsnm{Burdick}, \binits{N.Q.}},
\bauthor{\bsnm{Baumann}, \binits{K.}},
\bauthor{\bsnm{Tang}, \binits{Y.}},
\bauthor{\bsnm{Lu}, \binits{M.}},
\bauthor{\bsnm{Lev}, \binits{B.L.}}:
\batitle{Fermionic suppression of dipolar relaxation}.
\bjtitle{Phys. Rev. Lett.}
\bvolume{114},
\bfpage{023201}
(\byear{2015})
\end{barticle}
\endbibitem

\bibitem{Ticknor10}
\begin{barticle}
\bauthor{\bsnm{Ticknor}, \binits{C.}}:
\batitle{Quasi-two-dimensional dipolar scattering}.
\bjtitle{Phys. Rev. A}
\bvolume{81},
\bfpage{042708}
(\byear{2010})
\end{barticle}
\endbibitem

\bibitem{Condon47}
\begin{barticle}
\bauthor{\bsnm{Condon}, \binits{E.U.}}:
\batitle{The franck-condon principle and related topics}.
\bjtitle{American journal of physics}
\bvolume{15}(\bissue{5}),
\bfpage{365}--\blpage{374}
(\byear{1947})
\end{barticle}
\endbibitem

\bibitem{Hensler03}
\begin{barticle}
\bauthor{\bsnm{Hensler}, \binits{S.}},
\bauthor{\bsnm{Werner}, \binits{J.}},
\bauthor{\bsnm{Griesmaier}, \binits{A.}},
\bauthor{\bsnm{Schmidt}, \binits{P.}},
\bauthor{\bsnm{G{\"o}rlitz}, \binits{A.}},
\bauthor{\bsnm{Pfau}, \binits{T.}},
\bauthor{\bsnm{Giovanazzi}, \binits{S.}},
\bauthor{\bsnm{Rza{\.z}ewski}, \binits{K.}}:
\batitle{Dipolar relaxation in an ultra-cold gas of magnetically trapped
  chromium atoms}.
\bjtitle{Applied Physics B}
\bvolume{77}(\bissue{8}),
\bfpage{765}--\blpage{772}
(\byear{2003})
\end{barticle}
\endbibitem

\bibitem{Kagan85}
\begin{botherref}
\oauthor{\bsnm{Kagan}, \binits{Y.}},
\oauthor{\bsnm{Svistunov}, \binits{B.}},
\oauthor{\bsnm{Shlyapnikov}, \binits{G.}}:
Effect of bose condensation on inelastic processes in gases.
JETP Lett
\textbf{42}(4)
(1985)
\end{botherref}
\endbibitem

\bibitem{Burt97}
\begin{barticle}
\bauthor{\bsnm{Burt}, \binits{E.A.}},
\bauthor{\bsnm{Ghrist}, \binits{R.W.}},
\bauthor{\bsnm{Myatt}, \binits{C.J.}},
\bauthor{\bsnm{Holland}, \binits{M.J.}},
\bauthor{\bsnm{Cornell}, \binits{E.A.}},
\bauthor{\bsnm{Wieman}, \binits{C.E.}}:
\batitle{Coherence, correlations, and collisions: What one learns about
  bose-einstein condensates from their decay}.
\bjtitle{Phys. Rev. Lett.}
\bvolume{79},
\bfpage{337}--\blpage{340}
(\byear{1997})
\end{barticle}
\endbibitem

\bibitem{Ticknor09}
\begin{barticle}
\bauthor{\bsnm{Ticknor}, \binits{C.}}:
\batitle{Two-dimensional dipolar scattering}.
\bjtitle{Phys. Rev. A}
\bvolume{80},
\bfpage{052702}
(\byear{2009})
\end{barticle}
\endbibitem

\bibitem{Julienne10}
\begin{barticle}
\bauthor{\bsnm{Micheli}, \binits{A.}},
\bauthor{\bsnm{Idziaszek}, \binits{Z.}},
\bauthor{\bsnm{Pupillo}, \binits{G.}},
\bauthor{\bsnm{Baranov}, \binits{M.A.}},
\bauthor{\bsnm{Zoller}, \binits{P.}},
\bauthor{\bsnm{Julienne}, \binits{P.S.}}:
\batitle{Universal rates for reactive ultracold polar molecules in reduced
  dimensions}.
\bjtitle{Phys. Rev. Lett.}
\bvolume{105},
\bfpage{073202}
(\byear{2010})
\end{barticle}
\endbibitem

\bibitem{quemener2011dynamics}
\begin{barticle}
\bauthor{\bsnm{Qu{\'e}m{\'e}ner}, \binits{G.}},
\bauthor{\bsnm{Bohn}, \binits{J.L.}}:
\batitle{Dynamics of ultracold molecules in confined geometry and electric
  field}.
\bjtitle{Physical Review A}
\bvolume{83}(\bissue{1}),
\bfpage{012705}
(\byear{2011})
\end{barticle}
\endbibitem

\bibitem{Zoller07}
\begin{barticle}
\bauthor{\bsnm{Micheli}, \binits{A.}},
\bauthor{\bsnm{Pupillo}, \binits{G.}},
\bauthor{\bsnm{B\"uchler}, \binits{H.P.}},
\bauthor{\bsnm{Zoller}, \binits{P.}}:
\batitle{Cold polar molecules in two-dimensional traps: Tailoring interactions
  with external fields for novel quantum phases}.
\bjtitle{Phys. Rev. A}
\bvolume{76},
\bfpage{043604}
(\byear{2007})
\end{barticle}
\endbibitem

\bibitem{Browaeys20}
\begin{barticle}
\bauthor{\bsnm{Browaeys}, \binits{A.}},
\bauthor{\bsnm{Lahaye}, \binits{T.}}:
\batitle{Many-body physics with individually controlled rydberg atoms}.
\bjtitle{Nature Physics}
\bvolume{16}(\bissue{2}),
\bfpage{132}--\blpage{142}
(\byear{2020})
\end{barticle}
\endbibitem

\bibitem{Lin09}
\begin{barticle}
\bauthor{\bsnm{Lin}, \binits{Y.-J.}},
\bauthor{\bsnm{Compton}, \binits{R.L.}},
\bauthor{\bsnm{Jim{\'e}nez-Garc{\'\i}a}, \binits{K.}},
\bauthor{\bsnm{Porto}, \binits{J.V.}},
\bauthor{\bsnm{Spielman}, \binits{I.B.}}:
\batitle{Synthetic magnetic fields for ultracold neutral atoms}.
\bjtitle{Nature}
\bvolume{462}(\bissue{7273}),
\bfpage{628}--\blpage{632}
(\byear{2009})
\end{barticle}
\endbibitem

\bibitem{Lin11}
\begin{barticle}
\bauthor{\bsnm{Lin}, \binits{Y.-J.}},
\bauthor{\bsnm{Jim{\'e}nez-Garc{\'\i}a}, \binits{K.}},
\bauthor{\bsnm{Spielman}, \binits{I.B.}}:
\batitle{Spin--orbit-coupled bose--einstein condensates}.
\bjtitle{Nature}
\bvolume{471}(\bissue{7336}),
\bfpage{83}--\blpage{86}
(\byear{2011})
\end{barticle}
\endbibitem

\bibitem{burdick2016long}
\begin{barticle}
\bauthor{\bsnm{Burdick}, \binits{N.Q.}},
\bauthor{\bsnm{Tang}, \binits{Y.}},
\bauthor{\bsnm{Lev}, \binits{B.L.}}:
\batitle{Long-lived spin-orbit-coupled degenerate dipolar fermi gas}.
\bjtitle{Physical Review X}
\bvolume{6}(\bissue{3}),
\bfpage{031022}
(\byear{2016})
\end{barticle}
\endbibitem

\bibitem{Lunden20}
\begin{barticle}
\bauthor{\bsnm{Lunden}, \binits{W.}},
\bauthor{\bsnm{Du}, \binits{L.}},
\bauthor{\bsnm{Cantara}, \binits{M.}},
\bauthor{\bsnm{Barral}, \binits{P.}},
\bauthor{\bsnm{Jamison}, \binits{A.O.}},
\bauthor{\bsnm{Ketterle}, \binits{W.}}:
\batitle{Enhancing the capture velocity of a {Dy} magneto-optical trap with
  two-stage slowing}.
\bjtitle{Phys. Rev. A}
\bvolume{101},
\bfpage{063403}
(\byear{2020})
\end{barticle}
\endbibitem

\bibitem{lu2011spectroscopy}
\begin{barticle}
\bauthor{\bsnm{Lu}, \binits{M.}},
\bauthor{\bsnm{Youn}, \binits{S.H.}},
\bauthor{\bsnm{Lev}, \binits{B.L.}}:
\batitle{Spectroscopy of a narrow-line laser-cooling transition in atomic
  dysprosium}.
\bjtitle{Physical Review A}
\bvolume{83}(\bissue{1}),
\bfpage{012510}
(\byear{2011})
\end{barticle}
\endbibitem

\bibitem{Li2023}
\begin{botherref}
\oauthor{\bsnm{Du}, \binits{L.}},
\oauthor{\bsnm{Barral}, \binits{P.}},
\oauthor{\bsnm{Cantara}, \binits{M.}},
\oauthor{\bparticle{de} \bsnm{Hond}, \binits{J.}},
\oauthor{\bsnm{Lu}, \binits{Y.-K.}},
\oauthor{\bsnm{Ketterle}, \binits{W.}}:
Atomic physics on a 50 nm scale: Realization of a bilayer system of dipolar
  atoms.
arXiv
(2023)
\end{botherref}
\endbibitem

\bibitem{tang2015s}
\begin{barticle}
\bauthor{\bsnm{Tang}, \binits{Y.}},
\bauthor{\bsnm{Sykes}, \binits{A.}},
\bauthor{\bsnm{Burdick}, \binits{N.Q.}},
\bauthor{\bsnm{Bohn}, \binits{J.L.}},
\bauthor{\bsnm{Lev}, \binits{B.L.}}:
\batitle{s-wave scattering lengths of the strongly dipolar bosons dy 162 and dy
  164}.
\bjtitle{Physical Review A}
\bvolume{92}(\bissue{2}),
\bfpage{022703}
(\byear{2015})
\end{barticle}
\endbibitem

\end{thebibliography}

\pagebreak

\clearpage

\section*{Supplementary Information}

We present here our theoretical model to compute the expected rates shown in Fig.~\ref{fig:results}. The calculations use the formalism of Fermi's golden rule. For completeness, we also derive and discuss the Born approximation previously sketched in \cite{Pasquiou10} and used extensively in \cite{Li2023}. We comment on their equivalence and then discuss the pure-2D limit.

Both Fermi's golden rule and the Born approximation are perturbative approaches to quantum scattering. The former decomposes incoming and outgoing waves into many different channels, reducing the calculation effectively to 1D (i.e. the radial coordinate). The latter computes the perturbative impact of the dipolar relaxation Hamiltonian on an incoming plane wave, leaving the calculation in 2D. Both results are equivalent.  It is easier to generalize Fermi's golden rule for our approach, which uses initial wave functions modified by the dipolar elastic potential.  The Born approximation will be presented only for un-modified plane waves, whereas Fermi's golden rule formalism will be applied to both modified and un-modified wave functions.

\subsection*{Dipolar units}
For the following derivations, it is convenient to use dipolar units, indicated with a tilde. Dimensionless lengths are set with the dipolar length, such that $\tilde a_z = a_z/a_\rm{dd}$, wave functions become $\tilde \phi = a_\rm{dd}^{1/2}\phi$, momenta become $\tilde k_i = k_i a_\rm{dd}$ and energies are measured in units of dipolar energy $E_\rm{dd} = \frac{\hbar^2}{2\mu a_\rm{dd}^2}$, so $\tilde V_\rm{dd} = V_\rm{dd}/E_\rm{dd}$. The equation (\ref{eq:dipoleOperatorOnState}) becomes
\begin{equation}
    \hat V_\rm{dd}\ket{j_0} = \frac{2E_\rm{dd}}{\tilde r^3}\left[\left(1-3\bar{z}^2\right)\ket{j_0}
    - \frac{3\bar{z}\bar{r}_+}{J^{1/2}}\ket{j_1} - \frac{3\bar{r}_+^2}{2J}\ket{j_2}\right]
\end{equation}
and equation (\ref{eq:differentialEquation}) for an arbitrary harmonic oscillator channel $n$ becomes
\begin{equation}
    \left(-\frac{\d^2}{\d\tilde \rho^2} + \frac{m^2-1/4}{\tilde\rho^2} + \bra{n}\tilde V_{\rm{dd},\, 0}\ket{n} \right)\tilde \phi_{n,m} = \tilde k_i^2\tilde \phi_{n,m}.
    \label{eq:dimensionlessShieldedSchrodingerEq}
\end{equation}
The normalization condition reads: $\int_{0}^{\tilde L} \d\tilde\rho\,\tilde\phi_{n,m}(\tilde\rho)^2 = 1$ for a cylinder of radius $L = \tilde La_\rm{dd}$. Our model accounts for the modification of both incoming and outgoing wave functions by the dipolar interaction. The free radial wave function solution with momentum $k$ is $\tilde\phi_{n, m}^{(\mathrm{free})}(\tilde\rho) = \sqrt{\frac{\pi \tilde k\tilde\rho}{\tilde L}}J_{m}(\tilde k\tilde\rho)$, which does not depend on $n$.

\subsection*{Fermi's golden rule derivation}
Here we derive the expression for the 3D loss rate coefficient for the channel ${\ket{j_0}\rightarrow \ket{j_f}, \ket{0} \rightarrow \ket{n_f}}$:
\begin{equation*}
\beta_\rm{3D}^{j_f, n_f} = \frac{4\sqrt{\pi}}{\tilde{k}_i \tilde{k}_f}\tilde a_z\frac{E_\rm{dd}}{\hbar}a_\rm{dd}^3\left\vert\tilde{L}\int \d\tilde{z}\int_{0}^{\tilde{L}} \d\tilde{\rho}\,\tilde{\phi}_{n_f, j_f}(\tilde{\rho})\tilde{\chi}_{n_f}(\tilde{z})\tilde{V}_{\rm{dd}, j_f}(\tilde{\rho}, \tilde{z})\tilde{\chi}_{0}(\tilde{z})\tilde{\phi}_{0}(\tilde{\rho})\right\vert^2.
\end{equation*}
$\tilde\chi_{n}$ are the harmonic oscillator wave functions in the $z$ direction. $\tilde a_z$ is the harmonic oscillator length in units of the dipolar length.
\paragraph*{Fermi's golden rule} Fermi's golden rule conveniently expresses the decay rate from an initial state to a continuum having a certain density of states. It is natural to compute Fermi's golden rule with free incoming and outgoing plane waves, which we do at first. We then expand it into cylindrical waves, which we finally substitute for their shielded version solutions of equation~(\ref{eq:dimensionlessShieldedSchrodingerEq}).

The starting point is using equation~(\ref{eq:FermiGoldenRule}) to compute the decay rate $\Gamma_\rm{plane}^{j_f,n_f}$ of a pair of polarized atoms coming in a plane wave in the ground state of the harmonic oscillator, and outgoing in another asymptotic plane wave in the oscillator state $n_f$ with a spin state $\ket{j_f}$. The particles are assumed to be contained in a cylinder of length $L$ and by a harmonic oscillator potential in the $z$ direction.
\begin{equation*}
\hbar \Gamma_\rm{plane}^{j_f, n_f} = 2\pi\left\vert\bra{\Psi_{j_f, n_f}}\hat V_\rm{dd}\ket{\Psi_{j_0, 0}}\right\vert^2\rho_p(E_f),
\end{equation*}
with the total wave function $\ket{\Psi_{j, n}} = \ket*{\vec{k}}\otimes\ket{n}\otimes \ket{j}$ and
\begin{align*}
    \bra*{\vec{\rho}}\ket*{\vec{k}} &= \frac{e^{i\vec{k}\cdot\vec{\rho}}}{\sqrt{\pi L^2}}\\
    \bra{z}\ket{n} &= \frac{1}{\sqrt{2^n n!}}\left(\frac{1}{\pi a_z^2}\right)^{1/4}H_n(z/a_z)e^{-\frac{z^2}{2a_z^2}} = \chi_n(z)
\end{align*}

\paragraph*{Density of states}
In a two-dimensional box, the volumic density of states is $\frac{\mu}{2\pi\hbar^2}$. In a particular direction of angle $\d\theta_k$, the density of states is $\rho_p(k_f) = \frac{\mu L^2}{4\pi\hbar^2}\d\theta_k$. 
\paragraph*{Plane wave expansion}
To fully use the symmetries of the dipolar potential, one can expand the plane wave into spherical waves:
\begin{equation}
\label{eq:planeWaveExpansion}
\ket{\vec{k}} = \sqrt{\frac{2}{\pi k_i L}}\sum_{m = -\infty}^{+\infty}i^m e^{-im\theta_k} \ket{k, m},
\end{equation}
which have the following position representation:
\begin{equation}
    \label{eq:radialWaveFunction}
    \bra*{\vec{\rho}}\ket*{k,m} = \frac{e^{im\theta}}{\sqrt{2\pi}}\frac{\phi_{m}(\rho)}{\sqrt{\rho}}.
\end{equation}
Summing over all possible outgoing directions, the rate is
\begin{multline}
\label{eq:doubleSumFGR}
\hbar \Gamma_\rm{plane}^{j_f, n_f} = 2\pi\int\d\theta_f\Bigg\vert\sum_{m_i = -\infty}^{+\infty}\sum_{m_f = -\infty}^{+\infty}i^{m_i-m_f}e^{-im_i\theta_i+im_f\theta_f} \\ 
		\bra{k_f, m_f}\bra{n_f}\bra{j_f}\hat V_\rm{dd}\ket{j_0}\ket{0}\ket{k_i, m_i}\Bigg\vert^2\frac{2}{\pi k_i}\frac{2}{\pi k_f}\frac{\mu}{4\pi\hbar^2}.
\end{multline}
\paragraph*{Dipolar interaction} To simplify the problem we can look at the selection rules of the dipolar potential. The equation~(\ref{eq:dipoleOperator})
can be written
\begin{eqnarray*}
\hat{V}_{\rm{dd}} &=& \frac{\mu_0}{4\pi}(g_J\mu_B)^2\frac{\hat{\vec{J}}_1 \cdot \hat{\vec{J}}_2 - 3(\hat{\vec{J}}_1\cdot\vec{u}_r)(\hat{\vec{J}}_2\cdot\vec{u}_r)}{r^3} \\
&=& \frac{\mu_0}{4\pi}(g_J\mu_B)^2\frac{1}{r^3}\left\{ \right. \hat{J}_{1z}\hat{J}_{2z}\left(1-3\bar{z}^2\right) \\
                    & & + \frac{1}{2}\left(\hat{J}_{1+}\hat{J}_{2-}+\hat{J}_{1-}\hat{J}_{2+}\right) \\
                    & & - \frac{3}{2}\bar{z}\left[\hat{J}_{1z}\left(\hat{J}_{2+}\bar{r}_- + \hat{J}_{2-}\bar{r}_+\right) + \hat{J}_{2z}\left(\hat{J}_{1+}\bar{r}_- + \hat{J}_{1-}\bar{r}_+\right)\right] \\
                    & & - \frac{3}{4}\left(\hat{J}_{1+}\bar{r}_- + \hat{J}_{1-}\bar{r}_+\right)\left(\hat{J}_{2+}\bar{r}_- + \hat{J}_{2-}\bar{r}_+\right)\left.\right\}
\end{eqnarray*}
which gives equation~(\ref{eq:dipoleOperatorOnState})
\begin{eqnarray}
    \label{eq:dipolePotentialAppendix}
    \hat V_\rm{dd}\ket{j_0} =& \frac{2E_\rm{dd}}{\tilde r^3}\left[\left(1-3\bar{z}^2\right)\ket{j_0}
    - \frac{3\bar{z}\bar{r}_+}{J^{1/2}}\ket{j_1} - \frac{3\bar{r}_+^2}{2J}\ket{j_2}\right]. \\
    =& V_{\rm{dd},\, 0}\ket{j_0} + V_{\rm{dd},\, 1}\ket{j_1} + V_{\rm{dd},\, 2}\ket{j_2}
\end{eqnarray}
with $\bar{r}_{+} = \frac{\rho}{r}e^{i\theta}$ and $\bar z = z/r$. Hence $V_{\rm{dd},\, 0}$ is independent of $\theta$, $V_{\rm{dd},\, 1}$ proportional to $e^{i\theta}$ and $V_{\rm{dd},\, 2}$ to $e^{2i\theta}$. The $e^{im\theta}$ in equation~(\ref{eq:radialWaveFunction}) makes the matrix elements $\bra{k_f, m_f}\hat V_{\rm{dd},\, j_f}\ket{k_i, m_i}$ of the sum (\ref{eq:doubleSumFGR}) non-zero only if $m_f = m_i + j_f$. Furthermore, as $\bar{z}$ is anti-symmetric, and both $\chi_0$ and $\bar{r}_+$ are symmetric, $V_{\rm{dd},\, 1}$ can only promote to odd $n_f$ states and $V_{\rm{dd},\, 2}$ to even ones. This gives
\begin{equation}
\label{eq:FGR_sumPartialWaves}
\hbar \Gamma_\rm{plane}^{j_f, n_f} = \frac{2\mu}{\pi^2 k_i k_f\hbar^2}\int\d\theta_f\left\vert\sum_{m_i = -\infty}^{+\infty}e^{im_i\theta_f}\bra{k_f, m_i+j_f}\bra{n_f}V_{\rm{dd},\, j_f}\ket{0}\ket{k_i, m_i}\right\vert^2.
\end{equation}
\paragraph*{Symmetrization}
Since the atoms are bosons, the wave functions need to be symmetrized. All the spin states are already symmetrized. Each incoming and outgoing $\ket*{\vec{k}}$ state from (\ref{eq:planeWaveExpansion}) becomes $\frac{\ket*{\vec{k}}+\ket*{-\vec{k}}}{\sqrt{2}}$. It transforms the sum (\ref{eq:FGR_sumPartialWaves}) by multiplying the incoming and outgoing terms by $\sqrt{2}$ each and summing on even $m_i$ for bosons and odd for fermions. The density of states of the outgoing channels is divided by 2 to avoid double counting. Furthermore, due to the independence of the braket on $\theta_f$, the only terms in the sum giving a non-zero contribution after integrating over $\theta_f$ are the terms diagonal in $m_i$ coming from the modulus. This gives:
\begin{equation}
    \Gamma_{\text{sym}}^{j_f, n_f} = \frac{8\mu}{\pi k_i k_f\hbar^2}\sum_{\text{even } m_i}\left\vert \bra{k_f, m_i+j_f}\bra{n_f}V_{\rm{dd},\, j_f}\ket{0}\ket{k_i, m_i}\right\vert^2.
    \label{eq:FGR_allmi}
\end{equation}

\paragraph*{$s$-wave scattering}
Given our parameter range, we only keep the $m_i = 0$ channel, i.e. the $s$-wave channel.  This reflects that the relevant range of the dipolar potential is much smaller than the incoming De Broglie wavelength. It comes from the fact that $k_f \gg k_i$ for most of the magnetic fields (see in the "Comparing Born approximation and Fermi’s golden rule" section of the Supplementary Information for a discussion when the magnetic energy is comparable to the temperature). The outgoing wave function starts to oscillate at a distance $\sim \vert m_f\vert /k_f \sim 1/k_f$ which cuts out the integration at this distance (see Fig.~\ref{fig:plotWavefunctions_multiChannels}). Before that point, the free wave function $\phi_{m}(\rho)$ rises like a Bessel function in $\sqrt{\rho}(k\rho)^{\vert m\vert }$, and the dipolar potential goes as $1/\rho^3$. The integral goes then like $\int_0^{\vert m_f\vert /k_f}\rho\d\rho (k_f\rho)^{\vert m_f\vert }\frac{1}{\rho^3}(k_i\rho)^{\vert m_i\vert } \propto k_f^{\vert m_f\vert } k_i^{\vert m_i\vert }\frac{k_f^{-\vert m_i\vert -\vert m_f\vert +1}}{\vert m_i\vert +\vert m_f\vert -2} \propto k_f\left(\frac{k_i}{k_f}\right)^{\vert m_i\vert }\propto\sqrt{B}\left(\frac{k_B T}{B}\right)^{\vert m_i\vert /2}$, such that all the terms for which $m_i \neq 0$ are greatly suppressed as soon as the magnetic field energy is greater than the temperature (1$\mu$K, which is about 10\,mG). Therefore
\begin{equation*}
\hbar \Gamma_\rm{sym}^{j_f, n_f} = \frac{8\mu}{\pi k_i k_f\hbar^2}\left\vert\bra{k_f, m_f = j_f}\bra{n_f}V_{\rm{dd},\, j_f}\ket{0}\ket{k_i, 0}\right\vert^2.
\end{equation*}
\paragraph*{Wave function substitution}
We have expanded the plane wave in equation~(\ref{eq:planeWaveExpansion}) into cylindrical wave functions, solutions of the Schr\"odinger equation for a free particle, i.e. equation~(\ref{eq:dimensionlessShieldedSchrodingerEq}) without the dipolar interaction term. We now replace these free cylindrical wave functions $\phi_m$ by the solutions of the full Schr\"odinger equation with the dipolar interaction term $\phi_{n,m}$, which now depend on the harmonic oscillator channel $n$. The initial state is now the shielded wave function, and Fermi's golden rule describes its dipolar decay. This substitution is valid as the plane wave expansion~(\ref{eq:planeWaveExpansion}) still holds at large distance since the dipolar potential in $1/\rho^3$ decays faster than the centrifugal $1/\rho^2$ potential. 
\paragraph*{Calculation of $\beta_\rm{2D}$}
The rate  $\Gamma_{\text{sym}}$ is the probability per unit of time of the two bosons decaying when placed in a harmonic oscillator state $n=0$ in a cylindrical box of radius $L$. We are interested in the decay rate $\beta_\rm{2D}$ defined through $\frac{\d n_\rm{2D}}{\d t} = -\beta_\rm{2D} n_\rm{2D}^2$, which for a homogeneous gas gets integrated into $\frac{\d N}{\d t} = -\beta_{2D} \frac{N^2}{\pi L^2}$. For $N$ atoms homogeneously spread in the area, the differential equation sums the decay rate on all the possible pair combinations $N(N-1)/2$. For each event two atoms get lost, so $\frac{\d N}{\d t} = -2\Gamma\frac{N(N-1)}{2}\sim -\Gamma N^2$ for large $N$. It gives $\beta_\rm{2D} = \pi L^2\Gamma$: 
\begin{equation*}
\beta_\rm{2D}^{j_f, n_f} = \frac{8\mu}{k_i k_f\hbar^3}\left\vert L\int_{-\infty}^{+\infty}\d z\int_{0}^{L} \d\rho  \phi_{n_f, j_f}(\rho)\chi_{n_f}(z)V_{\rm{dd},\, j_f}(\rho, z)\chi_{0}(z)\phi_{0}(\rho)\right\vert^2.
\end{equation*}
The $\phi_{n, j}$ wave functions are either free wave functions like the light blue curve in Fig.~\ref{fig:plotWavefunctions_singleChannel}b or modified wave functions through equation~(\ref{eq:differentialEquation}) such as the blue and navy curves on the same figure.

\paragraph*{Calculation of $\beta_\rm{3D}$}
The final equation for $\beta_\rm{3D}$ follows from equation~(\ref{eq:beta2Dto3D}), so
\begin{equation}
\label{eq:3DdecayRate}
\beta_\rm{3D}^{j_f, n_f} = \frac{4\sqrt{\pi}}{\tilde{k}_i \tilde{k}_f}\tilde a_z\frac{E_\rm{dd}}{\hbar}a_\rm{dd}^3\left\vert\tilde{L}\int\int_{0}^{\tilde{L}} \d\tilde{z}\d\tilde{\rho} \, \tilde{\phi}_{n_f, j_f}(\tilde{\rho})\tilde{\chi}_{n_f}(\tilde{z})\tilde{V}_{\rm{dd}, j_f}(\tilde{\rho}, \tilde{z})\tilde{\chi}_{0}(\tilde{z})\tilde{\phi}_{0}(\tilde{\rho})\right\vert^2
\end{equation}
Note that the prefactor $E_\rm{dd}a_\rm{dd}^3/\hbar$ is useful to check the units but inconvenient to look at the scaling with the dipolar interaction. When the wave functions are not modified by the dipole-dipole potential they read: $\tilde \phi_{n, m}(\tilde \rho) = \sqrt{\frac{\pi \tilde k\tilde \rho}{\tilde L}}J_{m}(\tilde k\tilde \rho)$ and $\tilde \chi_n(\tilde z) = \frac{1}{\sqrt{2^n n!}}\left(\frac{1}{\pi \tilde a_z^2}\right)^{1/4}H_n(\tilde z/\tilde a_z)e^{-\frac{\tilde z^2}{2\tilde a_z^2}}$, which gives for the 2D rate:
\begin{multline*}
    \beta_\rm{2D}^\rm{free} =\frac{9}{2\pi J^2} \frac{\hbar}{\mu}a_\rm{dd}^2\Bigg\vert\int_{-\infty}^{+\infty}\int_{0}^{L}\d\xi \, 2\pi\rho \d{\rho} ~J_{j_f}(k_f \rho)\frac{\rho^2}{\left(\rho^2+\xi^2\tilde a_z^2\right)^{3/2}}J_{0}(k_i\rho) \\ 
    				\frac{1}{\sqrt{2^{n_f} n_f!}}H_{n_f}(\xi)e^{-\xi^2}\Bigg\vert^2
\end{multline*}
which explicitly shows the $a_\rm{dd}^2\propto (10\mu_B)^4$ scaling.

\paragraph*{Total decay rate}
The total rate is then the sum over all channels:
\begin{equation*}
\beta_{\rm{3D}} = \sum_{j_f, n_f}\beta_\rm{3D}^{j_f, n_f}.
\end{equation*}

\paragraph*{Momentum averaging}
The rate obtained depends on the incoming momentum $\tilde k_i$. The gas being thermal with many occupied states in the transverse direction, we integrate over momentum to obtain the average decay rate
\begin{equation*}
    \bar \beta_{\rm{3D}} = \frac{1}{\pi\kappa^2}\int_0^\infty 2\pi k\beta_\rm{3D}(k)e^{-k^2/\kappa^2}\d k
    \label{eq:beta3D_momentumAverage}
\end{equation*}
with $\kappa = \sqrt{k_BT/E_\rm{dd}}$. The results presented in Fig.~\ref{fig:results}b-c are the average momentum rates. But since this is rather computationally heavy and blurs the channel opening, we only present results computed at the mean momentum $\overline{\tilde k_i }= \kappa\sqrt{\pi}/2$ in the rest of the paper. For instance, the incoming energy of the wave functions in Fig.~\ref{fig:plotWavefunctions_singleChannel}a-b is $E_i =  E_\rm{dd} \kappa^2\pi/4$.

\subsection*{Van der Waals interaction}
Short-range van der Waals interactions exist on top of the dipolar ones. To get a sense of the sensitivity of our model to that contact interaction we also used simulated wave functions with a hard-core potential at $a_s~=~5.9$~nm, the effective background $s$-wave scattering length of dysprosium \cite{tang2015s}, while keeping the dipolar potential elsewhere. We put a node in the incoming and outgoing radial wave functions $\phi$ at this position, and integrated from this distance outward. This produced the lower bound of the shaded area in Fig.~\ref{fig:results}.

\subsection*{Born approximation}

\paragraph{Definitions}
The Born approximation is the standard way to describe scattering by a potential. The method in 2D has been described in \cite{Pasquiou10}, but it is lacking of an explicit final formula, which we would like to present here. We look for eigenstates of the full Hamiltonian:
\begin{equation}
\label{eq:BA_eigenvalue}
(E - \hat H_0)\ket{\Psi} = \hat V_\rm{dd}\ket{\Psi}
\end{equation}
with
\begin{equation}
\hat H_0 = H_{2D} + H_n + \hat H_B = -\frac{\hbar^2}{2\mu}\nabla_{2D}^2 + \hbar\omega_z\left(n +\frac{1}{2}\right) +\mu_B g_J B_z\left(\hat J_{z,1} + \hat J_{z,2} \right)
    \label{eq:H_0}
\end{equation}
and $\hat V_\rm{dd}$ defined through equation~(\ref{eq:dipolePotentialAppendix}). We define the Green operator $\hat G_0$ as $(E - \hat H_0 \pm i\eta)\hat G_0 = \mathbbm{1}$ and will eventually take the limit $\eta \to 0$. We omit the surface normalization coefficient in the wave function such that: $\bra*{\vec{\rho}}\ket*{\vec{k}} = e^{i\vec k\cdot\vec\rho}$; $\bra*{\vec{\rho}}\ket*{\vec{\rho}'} = \delta_{2D}(\vec\rho - \vec\rho')$ and $\bra*{\vec{k}}\ket*{\vec{k}'} = (2\pi)^2 \delta_{2D}(\vec k - \vec k')$ \footnote{It follows that:
\begin{align*}
\mathbbm{1}_\rm{2D} &= \frac{1}{(2\pi)^2}\int\d^2\vec{k}\dyad*{\vec k}{\vec k} = \int\d^2\vec{\rho}\dyad*{\vec \rho}{\vec \rho} \\
\Psi(\vec{\rho}) &= \bra*{\vec{\rho}}\ket*{\Psi} = \frac{1}{(2\pi)^2}\int\d^2\vec{k}\bra*{\vec\rho}\dyad*{\vec k}{\vec k}\ket*{\Psi} = \frac{1}{(2\pi)^2}\int\d^2\vec{k}e^{i\vec k \cdot \vec\rho}\psi(\vec k) \\
\psi(\vec k) &= \bra*{\vec k}\ket*{\Psi} = \int \d^2\vec\rho e^{-i\vec k \cdot\vec\rho}\Psi(\vec\rho)\\
\end{align*}
}.

\paragraph{Born approximation}

Following the definitions, $\ket{\Psi} = \hat G_0\hat V_\rm{dd}\ket{\Psi}$ is one solution of the eigenvalue equation~(\ref{eq:BA_eigenvalue}). If furthermore we have a solution $\ket{\Psi_0}$ to the equation $(E - \hat H_0)\ket{\Psi_0} = 0$ then $\ket{\Psi} = \ket{\Psi_0} + \hat G_0\hat V_\rm{dd}\ket{\Psi}$ is also a solution. The first Born approximation consists in computing the first-order part of the solution:
\begin{equation}
\label{eq:BA_psiFirstApproximation}
\ket{\Psi} = \ket{\Psi_0} + \hat G_0\hat V_\rm{dd}\ket{\Psi_0}
\end{equation}
Let us take $\ket{\Psi_0} = \ket*{\vec k_i}\otimes\ket{n = 0}\otimes\ket{j_0}$ which has the energy: $E = \frac{\hbar^2k_i^2}{2\mu} + \frac{1}{2}\hbar\omega_z + 16\mu_Bg_JB_z$. Note that we will later symmetrize the incoming state by replacing $\ket*{\vec k_i}$ by $\frac{\ket*{\vec k_i}+\ket*{-\vec k_i}}{\sqrt{2}}$, so it is good to keep track of the $e^{i\vec k_i\cdot\vec\rho}$ terms that will eventually become $\frac{e^{i\vec k_i\cdot\vec\rho} + e^{-i\vec k_i\cdot\vec\rho}}{\sqrt{2}}$.

\paragraph{Calculation of \texorpdfstring{$\psi_{n_f, j_f}(\vec k)$}{psi(k)}}
We decompose the problem into channels $(n_f, j_f)$. We eventually want the position representation of the scattered wave function but the Green operator has a simpler expression in momentum space. When projecting $\ket{\Psi}$ on $\bra{j_f}\bra{n_f}\bra*{\vec k}$ the first term in equation~(\ref{eq:BA_psiFirstApproximation}) disappears for ${(n_f, j_f) \neq (n_i = 0, j_0)}$. The matrix element to compute is thus
\begin{equation*}
\psi_{n_f, j_f}(\vec k) = \bra{j_f}\bra{n_f}\mel*{\vec k}{\hat G_0\hat V_\rm{dd}}{\vec k_i}\ket{0}\ket{j_0}
\end{equation*}

When acted on the left side, the Green operator passes through the states which are the eigenstates of $\hat H_0$ from equation~(\ref{eq:H_0}), giving
\begin{equation}
    \psi_{n_f, j_f}(\vec k) = \frac{1}{{\scriptstyle E-\frac{\hbar^2k^2}{2\mu} - \hbar\omega_z\left(n_f+\frac{1}{2}\right) - \mu_B g_J B_z\left(16-\Delta m\right) \pm i\eta}}\bra{j_f}\bra{n_f}\mel*{\vec k}{\hat V_\rm{dd}}{\vec k_i}\ket{0}\ket{j_0}
\end{equation}
with $\Delta m = 1, 2$ for channels $j_1, j_2$ respectively. By setting:
\begin{equation*}
\frac{\hbar^2k_f^2}{2\mu} = \frac{\hbar^2k_i^2}{2\mu} - \hbar\omega_z\Delta n + \mu_B g_J B_z \Delta m
\end{equation*}
with $\Delta n = n_f - n_i$, the denominator can be simplified to $\frac{\hbar^2}{2\mu}\left(k_f^2-k^2 \pm i\eta \right)$. We introduce the identity $\int \d^2\vec{\rho}\ket{\vec{\rho}}\bra{\vec{\rho}}$ to get
\begin{equation*}
    \psi_{n_f, j_f}(\vec k) = \frac{2\mu}{\hbar^2}\frac{1}{k_f^2 - k^2\pm i\eta}  \int\d^2\vec{\rho}\d z e^{-i(\vec k - \vec k_i)\cdot\vec\rho}\chi^*_{n_f}(z) V_{\rm{dd},\, j_f}(\vec\rho, z)\chi_0(z).
\end{equation*}
which uses the notation of equation~(\ref{eq:dipolePotentialAppendix}) for the potential.

\paragraph{Calculation of \texorpdfstring{$\Psi_{n_f, j_f}(\vec \rho)$}{psi(rho)}}
In position representation we have
\begin{multline*}
\Psi_{n_f, j_f}(\vec\rho) = \frac{1}{(2\pi)^2}\int\d^2\vec k e^{i\vec k\cdot\vec\rho}\psi_{n_f,j_f}(\vec k) \\
	\; = \frac{2\mu}{\hbar^2}\int\d^2\vec\rho_1\underbrace{\frac{1}{(2\pi)^2}\int\d^2\vec k\frac{e^{i\vec k\cdot\left(\vec\rho-\vec\rho_1\right)}}{k_f^2 - k^2 \pm i\eta}}_{I} e^{i\vec k_i\cdot\vec\rho_1}\int\d z \chi^*_{n_f}(z) V_{\rm{dd},\, j_f}(\vec\rho_1, z)\chi_0(z)
\end{multline*}
Computing the integral $I$ requires a few steps in Mathematica:
\begin{align*}
I &= \frac{1}{(2\pi)^2}\int_0^\infty k\d k\int_0^{2\pi}\d\theta_k\frac{e^{ik\vert \vec\rho-\vec\rho_1\vert \cos\theta_k}}{k_f^2-k^2 \pm i\eta} \\
 &= \frac{1}{(2\pi)^2}\int_0^\infty k\d k\frac{2\pi J_0\left(k\vert \vec\rho-\vec\rho_1\vert \right)}{k_f^2-k^2\pm i\eta} \\
 &= \frac{1}{(2\pi)^2}\left(-2\pi K_0\left(\sqrt{-k_f^2\mp i\eta}\vert \vec\rho-\vec\rho_1\vert \right)\right)\\
 &= -\frac{i}{4}H_0^{(1)}\left(\sqrt{k_f^2\pm i\eta} \vert \vec\rho-\vec\rho_1\vert \right) \xrightarrow[\eta \to 0]{} -\frac{i}{4}H_0^{(1)}\left(\sqrt{k_f^2} \vert \vec\rho-\vec\rho_1\vert \right),
\end{align*}
with $K_0$ being the Bessel K function with a complex argument and $H_0^{(1)}$ the Hankel function of the first kind. We only keep the $+$ solution from the square root in the Hankel function to have an outgoing flux and perform the far field expansion $H_0^{(1)}( k_f\vert \vec{\rho}-\vec{\rho}'\vert ) \simeq \sqrt{\frac{2}{\pi k_f\rho}}e^{ik_f\rho}e^{-ik_f\vec{u}_\rho\cdot\vec{\rho}'}e^{-i\pi/4}$, so by writing $\vec k_f = k_f\vec u_\rho$ we find
\begin{multline*}
\Psi_{n_f, j_f}(\vec\rho) = \frac{2\mu}{\hbar^2}\frac{-i}{4}\sqrt{\frac{2}{\pi k_f\rho}}e^{ik_f\rho}e^{-i\pi/4}\int\d^2\vec\rho_1 e^{-i\left(\vec k_f- \vec k_i\right)\cdot\vec\rho_1} \\
			\int\d z \chi^*_{n_f}(z) V_{\rm{dd},\, j_f}(\vec\rho_1, z)\chi_0(z).
\end{multline*}
To further simplify we first introduce the Fourier transform $\mathcal{H}_{n_f}$ of the harmonic oscillator wave functions product $\chi^*_{n_f}(z)\chi_{0}(z)$:
\begin{align*}
\chi^*_{n_f}(z)\chi_{0}(z) &= \frac{1}{2\pi}\int\d q_z e^{iq_z z}\mathcal{H}_{n_f}(q_z) = \frac{1}{2\pi}\int\d q_z e^{-iq_z z}\mathcal{H}_{n_f}(-q_z),
\end{align*}
and then define $\vec q = \vec k_f - \vec k_i + q_z\vec u_z$ and $\vec r_{1} = \vec \rho_1 + z\vec u_z$ which  gives
\begin{multline*}
\Psi_{n_f, j_f}(\vec\rho) = \frac{2\mu}{\hbar^2}\frac{-1}{4}\sqrt{\frac{2}{\pi k_f\rho}}e^{ik_f\rho}e^{+i\pi/4}\frac{1}{2\pi}\int\d q_z\mathcal{H}_{n_f}(-q_z)\\
\; \int\d^3\vec r_1 e^{-i\vec q\cdot\vec r_1} V_{\rm{dd},\, j_f}(\vec r_1).
\end{multline*}
We introduce the Fourier transform of the dipole-dipole interaction
\begin{equation*}
\mathcal{V}_{j_f} = \int\d^3\vec r_1 e^{-i\vec q\cdot\vec r_1} V_{\rm{dd},\, j_f}(\vec r_1),
\end{equation*}
and define the scattering amplitude $f$ such that
\begin{equation*}
\Psi_{n_f, j_f}(\vec\rho) = \frac{e^{ik_f\rho}}{\sqrt{\rho}}e^{i\pi/4}f(\vec k_f - \vec k_i, n_f, \Delta m).
\end{equation*}
Therefore
\begin{equation}
f(\vec k_f - \vec k_i, n_f, \Delta m) = \frac{\mu}{\hbar^2}\frac{-1}{2\sqrt{2}\pi^{3/2}}\frac{1}{\sqrt{k_f}}\int\d q_z\mathcal{H}_{n_f}(-q_z)\mathcal{V}_{\Delta m}(\vec k_f - \vec k_i, q_z).
\end{equation}
The Fourier transform is
\begin{multline}
    \mathcal{ \hat V}(\vec k, q_z) = \mu_0(Jg_J\mu_B)^2 \Bigg( \left(\hat k_z \hat k_z - 1\right)\ketbra{j_0}{j_0} + \\*
    \;  \frac{1}{J^{1/2}}\hat k_z\hat{k}_+\ketbra{j_1}{j_0} + \frac{1}{2J}\hat{k}_+\hat{k}_+\ketbra{j_2}{j_0}\Bigg)
\end{multline}
with $\bar k_+ = \frac{k_x + ik_y}{\sqrt{k^2 + q_z^2}}$ and $\bar k_z = \frac{q_z}{\sqrt{k^2 + q_z^2}}$
\paragraph{Symmetrization}
We carried all along a term $e^{i\vec k_i\cdot\vec\rho_1}$ which is $\Psi_0^{k_i}(\vec\rho_1)$. To symmetrize the bosonic wavefunction we simply change it to $\frac{e^{i\vec k_i\cdot\vec\rho_1}+e^{-i\vec k_i\cdot\vec\rho_1}}{\sqrt{2}}$ which leads to
\begin{equation*}
f_S(\vec k_f, \vec k_i, n_f, \Delta m) = \frac{1}{\sqrt{2}}\left(f(\vec k_f - \vec k_i, n_f, \Delta m) + f(\vec k_f + \vec k_i, n_f, \Delta m) \right).
\end{equation*}

\paragraph{Scattering cross section}
The flux is defined through the gradient of the wave function $\frac{\hbar}{\mu}\text{Re}\left[\frac{1}{i}\Psi^*\vec\nabla\Psi\right]$. $\vec{J}_i = \frac{\hbar\vec k_i}{\mu}$ is the incident current of $\Psi_0(\vec\rho) = e^{i\vec k_i\cdot \vec \rho}$. The main contribution to the outgoing wave current comes from its radial part as the other terms in the gradient fall off as $1/\rho^2$ instead of $1/\rho$: $\vec{J}_f \simeq \frac{\hbar\vec k_f}{\mu}\frac{1}{\rho}\vert f_S(\vec k_f, \vec k_i, n_f, \Delta m)\vert ^2$. The differential scattering cross section into an angle $\d\theta_f$ is
\begin{equation*}
\frac{\partial \sigma}{\partial \theta_f}\d\theta_f = \frac{k_f}{k_i}\vert f_S(\vec k_f, \vec k_i, n_f, \Delta m)\vert ^2.
\end{equation*}
So the total scattering cross section is
\begin{equation*}
\sigma(\vec k_i, n_f, \Delta m) = \frac{k_f}{k_i}\int\d\theta_f\vert f_S(\vec k_f, \vec k_i, n_f, \Delta m)\vert ^2.
\end{equation*}

\paragraph{\texorpdfstring{$\beta_\rm{2D}$}{beta2D} loss coefficient}
The loss coefficient is $\beta = \sigma v = \sigma\frac{\hbar k_i}{\mu}$. It still depends on the direction of $\vec k_i$. It is then averaged in all possible incoming directions: $\beta_{2D}(k_i, n_f, \Delta m) = \frac{1}{2\pi}\int \d\theta_i \beta_{2D}(\vec k_i, n_f, \Delta m)$ to give
\begin{multline*}
\beta_{2D}(k_i, n_f, \Delta m) = \frac{\mu}{\hbar^3}\frac{1}{32\pi^4}\int\d\theta_i\int\d\theta_f\bigg\vert \int\d q_z\mathcal{H}_{n_f}(-q_z)\\ \left(\mathcal{V}_{\Delta m}(\vec k_f - \vec k_i, q_z)+\mathcal{V}_{\Delta m}(\vec k_f + \vec k_i, q_z)\right)\bigg\vert ^2.
\end{multline*}

\subsection*{Comparing Born approximation and Fermi's golden rule}

\begin{figure}[h!]
    \centering   
    \includegraphics[width=0.6\columnwidth]{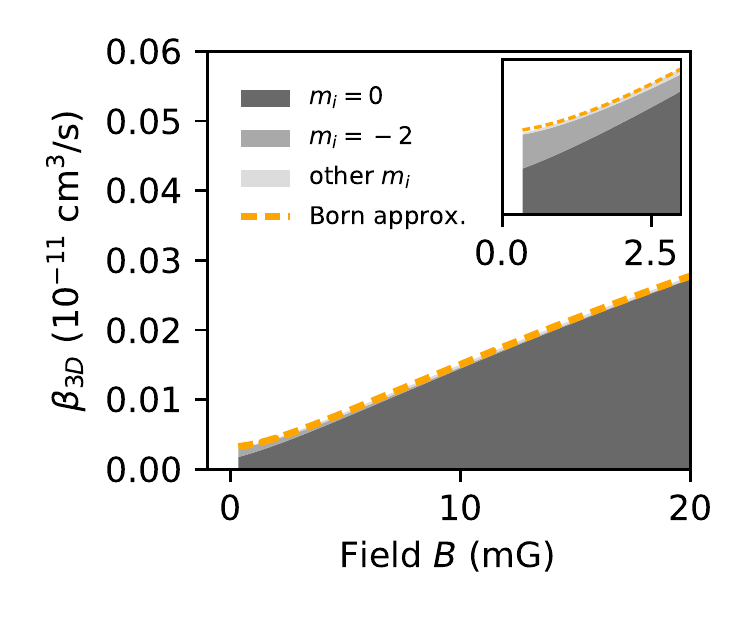}
    \caption{\textbf{Comparison between Fermi's golden rule and the Born approximation.} The regions shaded in gray are calculated with Fermi's golden rule from equation (\ref{eq:FGR_allmi}) with free wave functions, including multiple $m_i$ incoming partial waves. The orange line is the Born approximation. The insert zooms in on small magnetic fields.
    \label{fig:plotBornFGR}}
  \end{figure}

The Born approximation and Fermi's golden rule give the same results once all incoming $m_i$ partial waves are taken into account in the sum in equation~(\ref{eq:FGR_sumPartialWaves}). This is expected since they are both calculated in first-order perturbation theory.  It is not obvious from the final expressions but can be seen in Fig.~\ref{fig:plotBornFGR}. At high magnetic field, only the $m_i = 0$ channel is contributing since the rate scales as $k_f(k_i/k_f)^{\vert m_i\vert }$. However, at low enough magnetic field, the incoming and outgoing momenta are comparable and the other channels become as important, especially the $m_i = -2 \rightarrow 0$.

\subsection*{Pure-2D limit}
We mentioned in the main text that the decay rate in the pure-2D case is indefinitely suppressed when the magnetic field increases. The reason is that as the field increases, so does the outgoing momentum, and the relaxation becomes shorter-ranged. The inward region is shielded by the dipolar potential which repels the atoms. We offer here a derivation of the shielded rate in pure-2D. We first derive the un-shielded case. Then we use a known \cite{Ticknor09} zero-temperature shielded wave function and patch it at long-distance with a free low momentum wave function to compute the shielded rate. This will exhibit the spectacular behavior of a decreasing relaxation rate with an increasing magnetic field. We also show further suppression by going to lower temperature. We have not achieved this regime in our experimental setup, but we believe it is within reach with minor improvements, and think this theoretical treatment can provide insights into the system.

One starts with the equation (\ref{eq:3DdecayRate}) to write the 2D rate in the case of infinite transverse confinement ($\tilde\chi_0(z)= \delta(z)$). The only decay channel is $j_f = 2$ and
\begin{equation}
    \beta^\text{pure-2D} = \frac{4}{\tilde{k}_i \tilde{k}_f}\frac{E_\rm{dd}}{\hbar}a_\rm{dd}^2\left\vert\tilde{L}\int_{0}^{\tilde{L}} d\tilde{\rho} \tilde{\phi}_{f}(\tilde{\rho})\tilde{V}_{\rm{dd}, 2}(\tilde{\rho})\tilde{\phi}_{0}(\tilde{\rho})\right\vert^2
    \label{eq:FGR_pure2D}
\end{equation}
with $\tilde{V}_{\rm{dd}, 2}(\tilde{\rho}) = \left(\frac{-3}{2J}\right)\frac{2}{\rho^3}$.
\paragraph*{Free wave functions case}
If we ignore the shielding, the wave functions take a simple form $\tilde{\phi}_m(\tilde{\rho}) = \sqrt{\frac{\pi\tilde k\tilde \rho}{\tilde L}}J_{m}(\tilde k\tilde \rho)$. In the low-temperature limit, the incoming Bessel function goes like $J_{0}(\tilde k_i\tilde \rho)\simeq 1$. The integral $\int_0^{\infty}\d x\frac{J_2(kx)}{x^2}=\frac{k}{3}$ and
\begin{equation}
    \beta^\text{pure-2D}_\rm{free} = 4\pi^2\frac{1}{J^2}\frac{E_\rm{dd}}{\hbar}a_\rm{dd}^2\tilde{k}_f^2.
    \label{eq:betaPure2D_unShielded}
\end{equation}
So $\beta^\text{pure-2D}_\rm{free}\propto B$.

\paragraph*{Shielded wave functions case}
The result is radically different if we take into account the dipolar shielding potential. We will show two effects: the rate decreases both with increasing magnetic field and decreasing temperature. Lower temperature means greater shielding, and higher magnetic field shortens the range of the interaction which increases the shielding as well. To keep the model simple, we will only take the incoming wave function to be shielded since the outgoing one already experiences a centrifugal barrier. In pure-2D, the differential equation for the incoming wave function follows from equation (\ref{eq:differentialEquation}):
\begin{equation}
\label{eq:differentialEquation_2D}
    \left(-\frac{d^2}{d\tilde \rho^2} + \frac{-1/4}{\tilde\rho^2} + \frac{2}{\tilde\rho^3} \right)\tilde \phi = \tilde k_i^2\tilde \phi
\end{equation}
for which an analytical result is known from \cite{Ticknor09} but only at zero temperature ($\tilde k_i = 0$) and involves modified Bessel functions. At finite temperature and in the absence of the dipolar interaction term, the solution is known with regular Bessel functions. We therefore use the modified Bessel function solution at short-range until $\tilde \rho_0$ such that $2/\tilde \rho_0^3 = \tilde k_i^2$. We then patch this function with the finite temperature free solution with a phase shift $\delta$.
The incoming wave function then reads
\begin{align*}
    \tilde \phi_{\mathrm{in}}(\tilde \rho) =& \alpha\sqrt{\tilde \rho}K_0\left(\sqrt{\frac{8}{\tilde\rho}}\right) & \text{for $\rho<\rho_0$} \\
    =& \sqrt{\frac{\pi\tilde k\tilde \rho}{\tilde L}}\left(\cos(\delta) J_0(\tilde k\tilde \rho) - \sin(\delta) Y_0(\tilde k\tilde \rho)\right) & \text{for $\rho\geq\rho_0$}
\end{align*}
where $K_0$ is the modified Bessel function and $J_0$ and $Y_0$ the Bessel functions of the first and second kind. $\alpha$ is a normalization coefficient obtained by equating the wave functions and their derivative at the patching location $\tilde\rho = \tilde\rho_0$ in a box of length $\tilde L$, which gives
\begin{equation*}
    \alpha = \frac{\left(\cos(\delta) J_0(\tilde k_i\tilde \rho_0) - \sin(\delta) Y_0(\tilde k_i\tilde \rho_0)\right)\sqrt{\frac{\pi\tilde k_i\tilde\rho_0}{\tilde L}}}{\sqrt{\tilde\rho_0}K_0(\sqrt{\frac{8}{\tilde\rho_0}})}
\end{equation*}
with
\begin{equation}
    \delta = \arctan\left(\frac{rJ_0(\tilde k_i\tilde \rho_0)+J_1(\tilde k_i\tilde \rho_0)}{rY_0(\tilde k_i\tilde \rho_0)+Y_1(\tilde k_i\tilde \rho_0)}\right)
    \label{eq:phaseShift2D}
\end{equation}
and
\begin{equation*}
    r = \frac{1}{\tilde k_i}\left(\frac{\tilde\phi'(\tilde\rho_0)}{\tilde\phi(\tilde\rho_0)} - \frac{1}{2\tilde\rho_0}\right).
\end{equation*}

We now explore independently two limits: low temperature, and high magnetic field.

\paragraph*{Low temperature limit}
One can find the low temperature behavior of the incoming wave function. We expect the decay rate to be suppressed at low-temperature as the shielding increases. The $r$ coefficient can be rewritten:
\begin{equation*}
    r = \frac{K_1\left(\sqrt{\frac{8}{\tilde\rho_0}}\right)}{K_0\left(\sqrt{\frac{8}{\tilde\rho_0}}\right)},
\end{equation*}
which can be used to expand $\delta$ through equation~(\ref{eq:phaseShift2D})~\footnote{Note that equating this equation with the form of the phase shift in 2D for a hardcore potential $\tan \delta\simeq \frac{\pi}{2\left(\ln\left(\tilde k_i\tilde a/2\right)+\gamma \right)}$ allows to recover the universal dipolar scattering result from \cite{Ticknor09} that $\tilde a = 2e^{2\gamma}$.}:
\begin{equation*}
    \delta \xrightarrow[\tilde k_i \to 0]{} \mathrm{Arctan}\left(\frac{\pi}{2\left(3\gamma+\log(\tilde k_i)\right)}\right),
\end{equation*}
to find:
\begin{equation*}
    \alpha \xrightarrow[\tilde k_i \to 0]{} \frac{-2}{\log (\tilde k_i)}\sqrt{\frac{\pi\tilde k_i}{\tilde L}}.
\end{equation*}
It gives the following expression for the incoming wave function at low temperature:
\begin{equation*}
    \tilde \phi_{\mathrm{in}}(\tilde \rho) = \frac{-2}{\log (\tilde k_i)}\sqrt{\frac{\pi\tilde k_i\tilde\rho}{\tilde L}} K_0\left(\sqrt{\frac{8}{\tilde\rho}}\right).
\end{equation*}
This is reflected in the decay rate and $\beta^\mathrm{pure~2D}_\mathrm{shielded} \propto \left(1/\log(\tilde k_i)\right)^2$, which goes to zero at low temperature. This is very different from the free wave function case which had a finite limit even at zero temperature. The increase in the shielding factor with colder temperatures is shown in Fig.~\ref{fig:plotTheory}d.

\begin{figure}[h!]
    \centering
    \includegraphics[width=1\columnwidth]{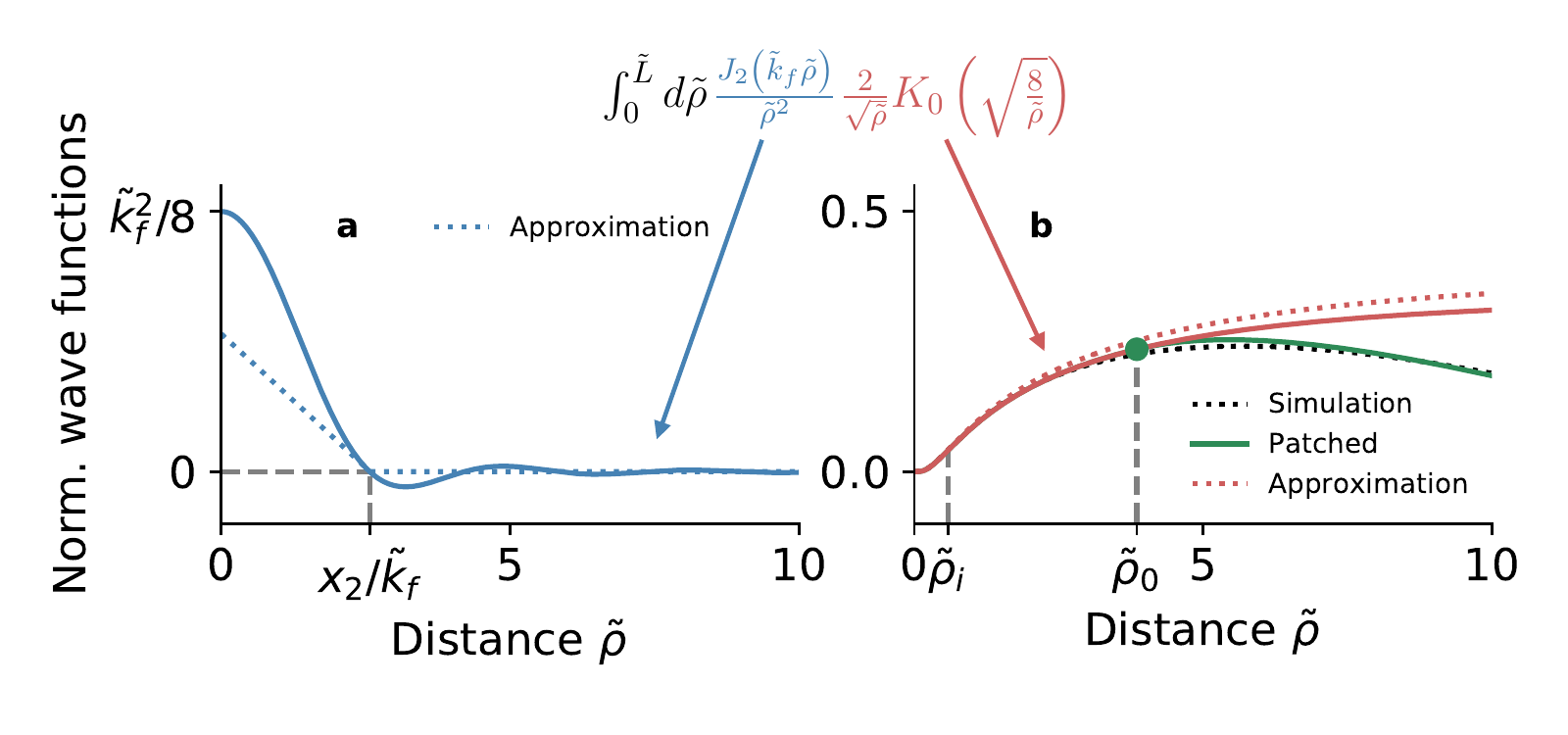}
    \caption{\textbf{Illustration of the approximations used to compute the shielded decay rate.} The left part of the integrand $J_2(\tilde k_f\tilde\rho)/\tilde\rho^2$ corresponding to the outgoing wave function, the potential and the surface element is plotted in solid blue. The outgoing wave function is cut off at the first zero in $x_2/\tilde k_f$ and then Taylor expanded at this point inward. The right part of the integrand corresponding to the incoming wave function is plotted in solid red. It is patched in $\tilde \rho_0$ to determine the normalization factor, and eventually expanded into a simple exponential function (dashed red). The total patched wave function (red and then green) is compared to the fully simulated wave function (dashed grey). Our model requires $\tilde\rho_i < x_2/\tilde k_f < \tilde\rho_0$.}
    \label{fig:plotAnalyticalFunctions}
\end{figure}

\paragraph*{Moderate magnetic field approximation}
Here we derive an approximate analytical formula for the integral in equation~(\ref{eq:FGR_pure2D}) with shielded wave functions. There is no simple expression for the integral of a Bessel $K$ function multiplied by a Bessel $J$ or $Y$ and the dipolar potential~\footnote{There is actually one involving Meijer G functions, but its doesn't bring in much insight compared to the approximations we make.}. We have done full numerical calculations (see below). However, we can find an approximate analytical result for moderate magnetic fields by approximating the Bessel functions by ones having known integrals. Fig.\,\ref{fig:plotAnalyticalFunctions} explains the different approximations we do to compute the integral. For high-enough magnetic field compared to the temperature\,\footnote{this corresponds to $x_2/\tilde k_f <\tilde\rho_0$ so $ x_2\hbar/(a_\rm{dd}\sqrt{4\mu\mu_B g_J B}) <\tilde\rho_0 = (2/\tilde k_i^2)^{1/3}$ which gives $B\simeq 0.3$\,G.}, the outgoing Bessel wave function oscillates and has its first zero before the patching point $\tilde\rho_0$ (see Fig.\ref{fig:plotAnalyticalFunctions}). We can then cut off the integral at this value as the remaining part is damped by the $1/\tilde\rho^3$ potential. This is valid only if the incoming wave function is not increasing too much after this zero, as it would compensate for the decrease due to the potential, which would result in the second lobe of the oscillation contributing more than the first one. The incoming wave function increases exponentially up to a certain distance $\tilde \rho_i$ which sets an upper bound on the magnetic field our model tolerates\,\footnote{the upper bound is $x_2/\tilde k_f >\tilde\rho_i\simeq 0.58$ which corresponds to $B\simeq 13$\,G. This is on the conservative side as the model can tolerate fields up to 50\,G.}. Defining $\tilde D = x_2/\tilde k_f < \tilde\rho_0$ with $x_2\simeq 5.13$ the first zero of $J_2(x)$, the decay rate from equation~(\ref{eq:FGR_pure2D}) is
\begin{align*}
    \beta^\text{pure-2D}_\mathrm{shielded} &= 
     \frac{4}{\tilde{k}_i \tilde{k}_f}\frac{E_\rm{dd}}{\hbar}a_\rm{dd}^2\left\vert\tilde{L}\int_{0}^{\tilde{D}} d\tilde{\rho} \sqrt{\frac{\pi\tilde k_f\tilde \rho}{\tilde L}}J_{2}(\tilde k_f\tilde \rho)\frac{2}{\tilde{\rho}^3}\alpha\sqrt{\tilde \rho}K_0\left(\sqrt{\frac{8}{\tilde\rho}}\right)
\right\vert^2 \\
    &= \frac{36\pi^2}{J^2}\frac{E_\rm{dd}}{\hbar}a_\rm{dd}^2\left(\frac{\cos(\delta) J_0(\tilde k_i\tilde \rho_0) - \sin(\delta) Y_0(\tilde k_i\tilde \rho_0)}{K_0(\sqrt{\frac{8}{\tilde\rho_0}})}\right)^2 \times \\
    & \; \; \;  \; \; \;  \; \; \; \left\vert\int_{0}^{\tilde{D}} d\tilde{\rho} J_{2}(\tilde k_f\tilde \rho)\tilde{\rho}^{-2}K_0\left(\sqrt{\frac{8}{\tilde\rho}}\right)
\right\vert^2 .
\end{align*}

\begin{figure}
    \centering
    \includegraphics[width=1\columnwidth]{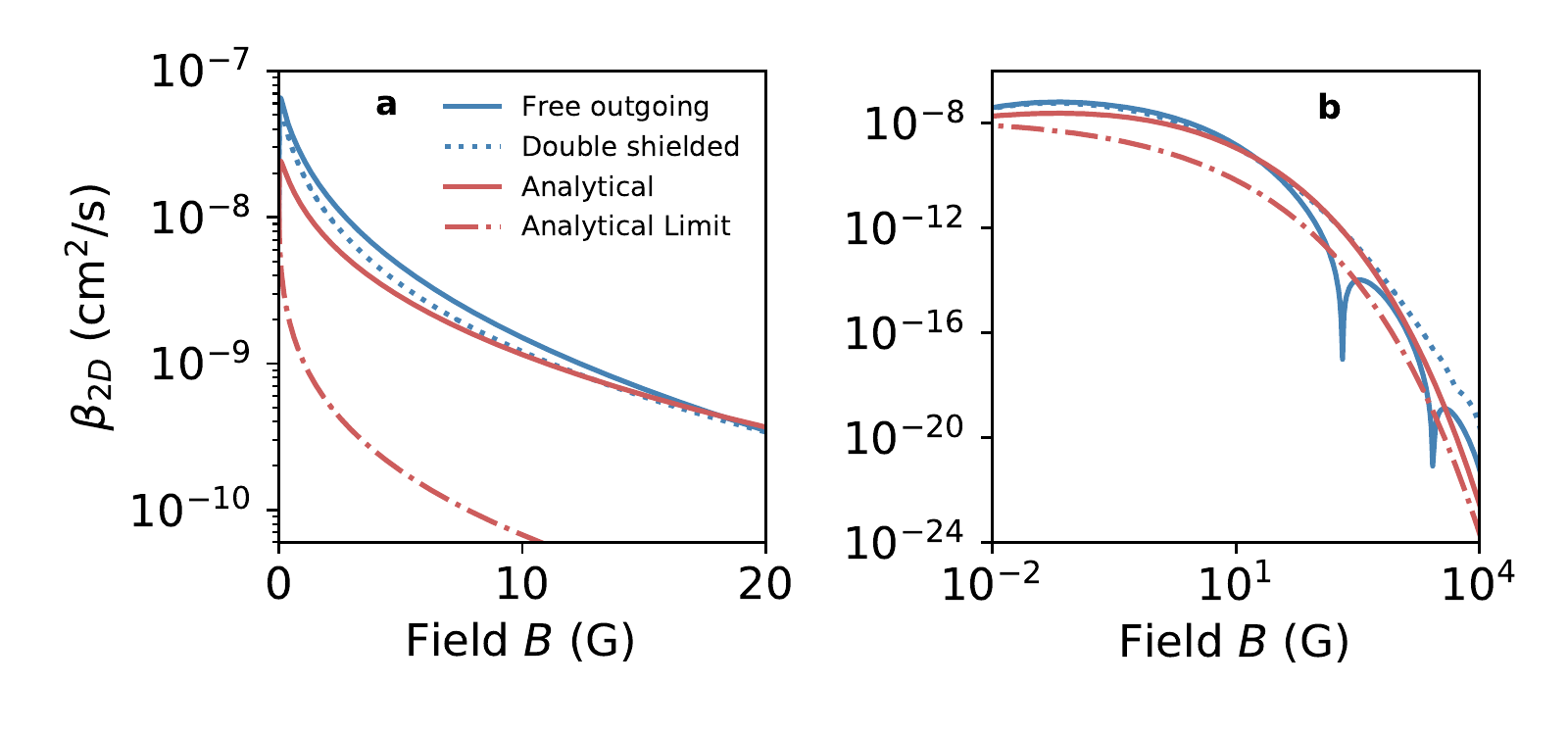}
    \caption{\textbf{Dipolar decay rates in two dimensions.} Shown are simulated and analytical decay rate coefficients for two magnetic field ranges. The blue curves represent the rate with simulated wave functions taking into account the dipolar repulsion for both the incoming and outgoing channels (dashed) or only on the incoming one (blue). The red curves present analytical results where the shielding is only accounted in the incoming wave function. The solid line is equation (\ref{eq:betaPure2D_Shielded}) and the dashed one is its high-field limit, equation (\ref{eq:betaPure2D_ShieldedLimit}).}  
    \label{fig:plotAnalyticalRates}
\end{figure}

The patching is only used to determine the normalization factor of $\tilde\phi_f$. As we mentioned, we further simplify the functional form of the wave function as illustrated in Fig.~\ref{fig:plotAnalyticalFunctions}. We use a Taylor expansion in $\tilde\rho=x_2/\tilde k_f$ of the Bessel $J$ function
\begin{equation*}
    \frac{J_2(\tilde k_f\tilde\rho)}{\tilde\rho^2}\simeq\tilde k_f^3\left(\frac{J_1(x_2)-J_3(x_2)}{2x_2^2}\right)\left(\tilde\rho-\frac{x_2}{\tilde k_f}\right),
\end{equation*}
and approximate the Bessel K function in $\tilde\rho = 0$ by
\begin{equation*}
    K_0\left(\sqrt{\frac{8}{\tilde\rho}}\right)\simeq \exp\left(-\sqrt{\frac{8}{\tilde\rho}}\right)\frac{\sqrt{\pi}\tilde\rho^{1/4}}{2\cdot 2^{1/4}}
\end{equation*}
The integral
\begin{equation*}
    \int_{0}^{\tilde{D}} d\tilde{\rho} \left(1-\frac{\tilde\rho}{\tilde D}\right)\tilde\rho^{1/4}e^{-\sqrt{\frac{8}{\tilde\rho}}} = \frac{16}{945}f\left(\tilde D\right)
\end{equation*}
with
\begin{multline*}
    f\left(\tilde D\right) = \frac{1}{\tilde D}\left[e^{-\sqrt{\frac{8}{\tilde D}}}\tilde D^{1/4}\left(-256+32\sqrt{2}\tilde D^{1/2}+480\tilde D-48\sqrt{2}\tilde D^{3/2}+21\tilde D^2\right)\right.\\ \left. -8\cdot2^{3/4}\sqrt{\pi}\left(-32+63\tilde D\right)\text{erfc}\left( \frac{2^{3/4}}{\tilde D^{1/4}}\right)\right]
\end{multline*}
gives overall
\begin{multline}
    \beta^\text{pure-2D}_\mathrm{shielded} = \frac{256\pi^3}{99225\sqrt{2}}\left(\frac{J_1(x_2)-J_3(x_2)}{2x_2}\right)^2 \times\\ 
    	\; \left(\frac{\cos(\delta) J_0(\tilde k_i\tilde \rho_0) - \sin(\delta) Y_0(\tilde k_i\tilde \rho_0)}{K_0(\sqrt{\frac{8}{\tilde\rho_0}})}\right)^2 \frac{1}{J^2}\frac{E_\rm{dd}}{\hbar}a_\rm{dd}^2\tilde k_f^4f^2\left(\frac{x_2}{\tilde k_f}\right).
    \label{eq:betaPure2D_Shielded}
\end{multline}
The agreement of this analytical expression with the numerical integration is presented in Fig.\,\ref{fig:plotAnalyticalRates}.

\paragraph*{High field limit}
The complexity of $f$ makes difficult to grasp the behavior of the decay rate $\beta$.  It is possible to do an expansion at high-magnetic fields of $f$ which greatly simplifies the equation. However this expansion becomes valid only for fields of several thousands of Gauss, which invalidates the initial assumptions that $x_2/\tilde k_f$ is larger than the limit $\tilde \rho_i$ of the exponentially suppressed region of the incoming wave function. Nonetheless it describes well the overall behavior of the curve, and brings some insight into the suppression of the decay rate. In the large field limit, when setting the numerical and normalization prefactor $\kappa$ to be:
\begin{equation*}
    \kappa = \frac{8\pi^3}{105}\left(\frac{J_1(x_2)-J_3(x_2)}{2x_2}\right)^2\left(\frac{\cos(\delta) J_0(\tilde k_i\tilde \rho_0) - \sin(\delta) Y_0(\tilde k_i\tilde \rho_0)}{K_0(\sqrt{\frac{8}{\tilde\rho_0}})}\right)^2x_2^{7/4}
\end{equation*}
one obtains
\begin{equation}
    \beta^\text{pure-2D}_ \mathrm{shielded} = \kappa\frac{1}{J^2}\frac{E_\rm{dd}}{\hbar}a_\rm{dd}^2\tilde k_f^{1/4}\exp\left(-2\sqrt{\frac{8\tilde k_f}{x_2}}\right),
    \label{eq:betaPure2D_ShieldedLimit}
\end{equation}
or $\beta^\text{pure-2D}_ \mathrm{shielded} \propto B^{1/8}\exp\left(-\xi B^{1/4}\right)$, which is a radically different behavior from the free wave function case where the rate is increasing linearly with the magnetic field. Fig.~\ref{fig:plotAnalyticalRates} shows this exponential suppression. The suppression factor can be made arbitrarily large. The overall shape of the curve is correctly reproduced by our analytical formulas~(\ref{eq:betaPure2D_Shielded})~and~(\ref{eq:betaPure2D_ShieldedLimit}) even at high fields, although our approximation does not fully capture the precise amplitude of nor the zeros in the decay rate that appear when $x_2/\tilde k_f$ becomes $< \tilde\rho_i$.

\paragraph*{Classical turning point}
We have explained in the main text how the Franck-Condon principle predicts spin-flips to occur at the classical turning point of the outgoing wave function. The low temperature shielded situation we just presented provides a counter-example where the incoming wave function also has a classical turning point that needs to be taken into account. When the field is sufficiently high, the outgoing wave function oscillates multiple times in the suppressed region of the incoming wave function. Therefore the integrand gets contributions from multiple oscillations, not just the first one.

\subsection*{Integrand behavior}
\label{appendix:multiChannels}
\begin{figure}
    \centering
    \includegraphics[width=0.6\columnwidth]{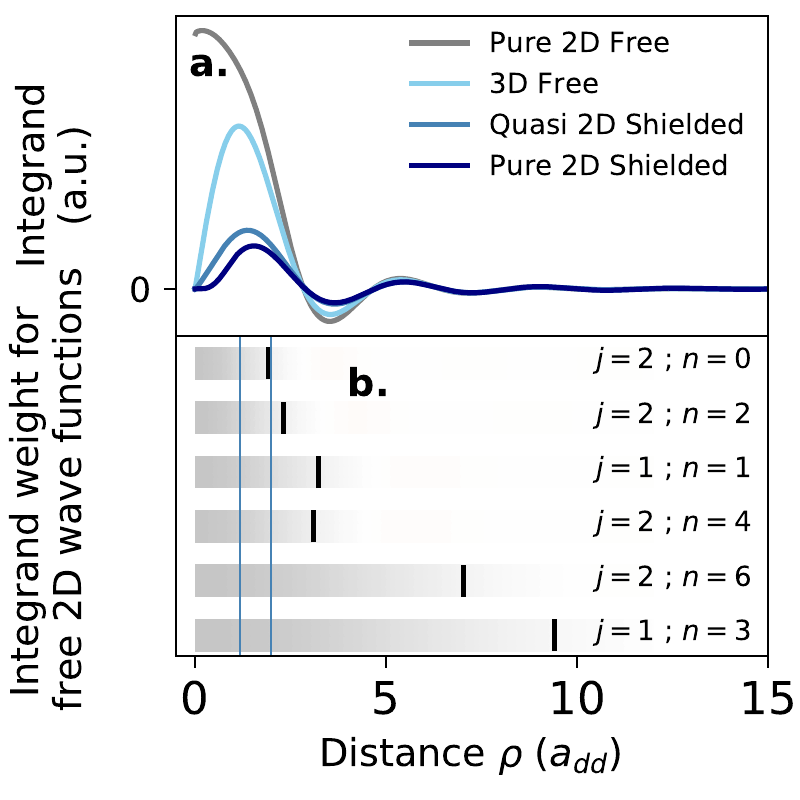}
    \caption{\textbf{Integrand of Fermi's golden rule.} (a.) The blue curves are the same as in Fig.~\ref{fig:plotWavefunctions_singleChannel} and come from equation~(\ref{eq:3DdecayRate}). The gray curve, which has been scaled down at short range, represents the fictitious case of a pure-2D geometry using free wave functions. (b.) Amplitude of the same integrand (for free wave functions) shown in grayscale. Multiple $n$-channels for single $j=1$ and double $j=2$ spin flip are presented. The top bar is the same as the gray curve in (a.). The vertical blue lines represent the interval where the dipolar interaction dominates over the other energy terms in equation~(\ref{eq:differentialEquation}) for the quasi-2D case. The vertical black bars are placed at $\sqrt{12}/k_f$, where $k_f$ is the outgoing momentum.}  
    \label{fig:plotWavefunctions_multiChannels}
  \end{figure}

The dipolar relaxation rate in three dimensions has been calculated elsewhere \cite{Hensler03, Pasquiou10, Lev15}. In reference~\cite{Pasquiou10}, the authors show that the rate gets its main contribution from the region around the classical turning point of the particles in the exit channel. This can be understood by the fact that the integrand in equation~(\ref{eq:FermiGoldenRule}) is the product of a flat incoming wave function, a $1/r^3$ potential, a spherical Bessel function that increases as $(k_fr)^2$ and a volume element $4\pi r^2$. The integrand goes like $r$ and increases up to $r \approx 1/k_f$ where the Bessel function starts to oscillate. In two dimensions the story is different, as shown in Fig.~\ref{fig:plotWavefunctions_multiChannels}a. The volume element being $2\pi\rho$, the integrand for free wave functions in pure-2D becomes flat and the contribution to the relaxation rate is homogeneous up to $\sim 1/k_f$. Therefore, there is a large contribution from the inner region, which is classically forbidden.

Going from the fictitious case of free wave functions in 2D to the real quasi-2D case with shielded wave functions has two effects. First, the spin-flip potential gets averaged along $z$, which reduces the short-range contribution of the integrand. Indeed, in quasi-2D $\langle \frac{\bar{r}_+^2}{ r^3}\rangle_{n = 0} \propto 1/a_z\rho^2$, which makes the integrand go to $0$ as $\rho \rightarrow 0$ (see Fig.~\ref{fig:plotWavefunctions_multiChannels}a, light blue). Then the shielding reduces the short-range amplitude even further (see Fig.~\ref{fig:plotWavefunctions_multiChannels}a, steel blue).

The two vertical blue lines in Fig.~\ref{fig:plotWavefunctions_multiChannels}b indicate the region of space where the dipolar interaction dominates over the centrifugal and the kinetic energy terms. This could also be inferred from Fig.~\ref{fig:plotWavefunctions_singleChannel}a by looking at when the dipolar interaction contribution to the quasi-2D blue curve is bigger than the absolute value of the centrifugal light blue curve and the incoming kinetic energy. The shielding mainly occurs in this region of space and one can only hope to see a reduction of dipolar relaxation from this region inward.

Therefore, the higher the outgoing momentum is, the shorter the range of the interaction is, which increases the shielding factor. This can be seen in Fig.~\ref{fig:plotTheory}c as the shielding factor increases with the magnetic field. Similarly, the excitation of axial motion reduces the final momentum in the radial direction and therefore moves the Franck-Condon point further out. Fig.~\ref{fig:plotWavefunctions_multiChannels}b shows that the integrand contributes far outside the shielded inner region for channels with smaller outgoing radial momentum.

\subsection*{Magnetic field scaling in 3D}
We mentioned in the main text that the decay rate for bosons scales as $\sqrt{B}$. This result can be understood with Fermi's golden rule. The outgoing wave function $\psi$ in 3D is an $l=2$ spherical Bessel function. Its normalization condition in a sphere of radius $L$ gives $\psi\propto k_f/\sqrt{L}j_2(k_f r)$. It therefore rises as $k_f^3r^2$ for small $r$ before it starts oscillating at $1/k_f$. By integrating the product of the incoming flat wave function, the outgoing one, the volume element $4\pi r^2$ and the potential $1/r^3$ between 0 and $1/k_f$, one gets a matrix element proportional to $k_f$. As we are considering spherical waves indexed by $k_f$, we use the one-dimensional density of states $\propto 1/k_f$. This overall gives
\begin{equation*}
\Gamma_\rm{3D} \propto k_f \propto \sqrt{B}.
\end{equation*}
Getting this same result by summing all the contributions from the 2D channels is more complicated as the harmonic oscillator's wave functions play a role, but one can see in Fig.~\ref{fig:results}b that the free 2D curve eventually meets the 3D one.
\end{document}